%% file: ris_isac.tex
\newif\ifsingle
\singletrue 

\newif\ifFullVersion

\ifsingle
\documentclass[11pt,draftclsnofoot, onecolumn]{IEEEtran}		
\else		
\documentclass[10pt,final, twocolumn]{IEEEtran}
\fi

\usepackage{bm}
\usepackage{balance}
\usepackage{times}
\usepackage{amsmath,dsfont}
\usepackage{amssymb}
\usepackage{epsfig,verbatim} 
\usepackage{setspace}
\usepackage{color}
\usepackage{cite}
\usepackage{epstopdf}
\usepackage{subfigure}
\usepackage{graphics}
\usepackage{accents}
\usepackage{acronym}
\usepackage[bookmarks,colorlinks]{hyperref}
\usepackage{ulem}
\usepackage{mathtools}
\usepackage{algorithm}
\usepackage{algorithmic}
\usepackage{enumitem}
\usepackage[most]{tcolorbox}
\usepackage{cuted}
\tcbuselibrary{skins}
\usepackage{multicol,lipsum}

\newcommand{\myVec}[1]{{\boldsymbol{#1}}}
\newcommand{\myMat}[1]{{\boldsymbol{#1}}}

 \newcommand{\NTx}{L_{\rm T}}
 \newcommand{\NRxRad}{L_{\rm S}}
 \newcommand{\NRxCom}{L_{\rm C}}
 \newcommand{\NRis}{N}
 \newcommand{\SigTx}{\myVec{x}}
 \newcommand{\SigRxRad}{\myVec{y}_s}
 \newcommand{\SigRxCom}{\myVec{y}_c}
 \newcommand{\ChRad}{\myVec{h}_r}
 \newcommand{\ChCom}{\myVec{h}_c}

\definecolor{mypurple}{rgb}{0.910, 0.910, 0.969}
\definecolor{myblue}{rgb}{0.122, 0.435, 0.698}

\acrodef{ris}[RIS]{reconfigurable intelligent surface} 
\acrodef{em}[EM]{electromagnetic} 
\acrodef{rf}[RF]{radio frequency} 
\acrodef{bs}[BS]{base station} 
\acrodef{ue}[UE]{user equipment} 
\acrodef{isac}[ISAC]{integrated sensing and communications} 
\acrodef{mimo}[MIMO]{multiple-input multiple-output} 
\acrodef{dfrc}[DFRC]{dual-function radar-communications}
\acrodef{los}[LoS]{line of sight} 
\acrodef{nlos}[NLoS]{non-line-of-sight}
\acrodef{aoa}[AOA]{angle of arrival}
\acrodef{doa}[DOA]{direction of arrival}
\acrodef{ofdm}[OFDM]{orthogonal frequency division multiplexing}
\acrodef{snc}[S\&C]{sensing and communications}
\acrodef{ula}[ULA]{uniform linear array} 
\acrodef{crb}[CRB]{Cram\'er-Rao bound}
\acrodef{qos}[QoS]{quality of service}
\acrodef{sinr}[SINR]{signal-to-interference-plus-noise ratio} 
\input{macros}

\IEEEoverridecommandlockouts

\title{Integrated Sensing and Communications with Reconfigurable Intelligent Surfaces}
\author{
{Sundeep Prabhakar Chepuri,~\IEEEmembership{Member,~IEEE}, Nir Shlezinger,~\IEEEmembership{Member,~IEEE}, \\Fan Liu,~\IEEEmembership{Member,~IEEE}, George C. Alexandropoulos,~\IEEEmembership{Senior~Member,~IEEE},  \\Stefano Buzzi,~\IEEEmembership{Senior~Member,~IEEE}, and Yonina C. Eldar,~\IEEEmembership{Fellow,~IEEE}}
\thanks{
		S. P. Chepuri is with the Department of ECE, Indian Institute of Science (IISc), Bengaluru, India (email: spchepuri@iisc.ac.in).
		 N. Shlezinger is with the School of ECE, Ben-Gurion University of the Negev, Beer-Sheva, Israel (e-mail: nirshl@bgu.ac.il).
		 F. Liu is with the Department of Electrical and Electronic Engineering, Southern University of Science and Technology (SUSTech), Shenzhen, China (email: liuf6@sustech.edu.cn).
		 G.~C.~Alexandropoulos is with the Department of Informatics and Telecommunications, National and Kapodistrian University of Athens, 15784 Athens, Greece and the Technology Innovation Institute, 9639 Masdar City, Abu Dhabi, United Arab Emirates (e-mail:  alexandg@di.uoa.gr). 
		 S. Buzzi is with the University of Cassino and Lazio Meridionale, Cassino, Italy (email: buzzi@unicas.it). He is also affiliated with Politecnico di Milano, Milano, Italy and with Consorzio Nazionale Interuniversitario per le Telecomunicazioni (CNIT), Parma, Italy.
		 Y. C. Eldar is with the Math and CS Faculty, Weizmann Institute of Science, Rehovot, Israel (e-mail:  yonina.eldar@weizmann.ac.il). 
	}
	\thanks{S.P. Chepuri is supported by the Next Generation Wireless Research
and Standardization on 5G and Beyond project (grant number 13(44)/2020-CC\&BT), MeitY, Govt. of India. G. C. Alexandropoulos and S. Buzzi are supported by the EU H2020 RISE-6G project (grant number 101017011) and the EU H2020 MSCA-ITN project META WIRELESS (grant number 956256), respectively.
}
} 

\begin{document}

\maketitle
\pagestyle{plain}
\thispagestyle{plain}
	

\begin{abstract}
\label{sec:abstract}
Integrated sensing and communications (ISAC) are envisioned to be an integral part of future wireless networks, especially when operating at the millimeter-wave (mmWave) and terahertz (THz) frequency bands. However, establishing wireless connections at these high frequencies is quite challenging, mainly due to the penetrating pathloss that prevents reliable communication and sensing. Another emerging technology for next-generation wireless systems is \acp{ris}, which are capable of modifying harsh propagation environments. \acp{ris} are the focus of growing research and industrial attention, bringing forth the vision of smart and programmable signal propagation environments. In this article, we provide a tutorial-style overview of the applications and benefits of \acp{ris} for sensing functionalities in general, and for ISAC systems in particular. We highlight the potential advantages when fusing these two emerging technologies, and identify for the first time that: \textit{i}) joint \acl{snc} designs are most beneficial when the channels referring to these operations are coupled, and that \textit{ii}) \acp{ris} offer means for controlling this beneficial coupling. The usefulness of \ac{ris}-aided ISAC goes beyond the individual obvious gains of each of these technologies in both performance and power efficiency. We also discuss the main signal processing challenges and future research directions which arise from the fusion of these two emerging technologies. 
\end{abstract}

\acresetall

\section{Introduction}
\label{sec:introduction}
 Recent years have witnessed growing research and industrial attention in \acp{ris} \cite{RIS_standardization}. An \ac{ris} is an array of elements with programmable scattering properties \cite{huang2019reconfigurable}. While this definition accommodates a broad range of technologies, the common treatment of \acp{ris} considers two-dimensional arrays whose elements  can be tuned independently to generate desirable reflection patterns in a nearly passive fashion, without utilizing active \ac{rf}  chains to process the impinging signals \cite{di2019smart}. The core benefits of \acp{ris} stem from their ability to shape the propagation profile of information-bearing \ac{em} waves in a flexible, low-cost, and energy efficient manner \cite{risTUTORIAL2020}.

 One of the most common applications of \acp{ris} is in wireless communications. This metamaterial-based technology brings forth the vision of smart programmable environments \cite{RISE6G_COMMAG}, where \acp{ris} are expected to improve coverage, energy efficiency, reliability, and EM field exposure of wireless communications \cite{chowdhury20206g}. The role of \acp{ris} in future wireless communications in modifying harsh propagation environment and establishing reliable links for communication via \ac{mimo} systems is widely studied in the literature (see, e.g., the recent survey papers \cite{risTUTORIAL2020, WavePropTCCN,CE_overview_2022,liu2021reconfigurable,RIS_6G_tutorial}). Consequently, they are expected to play a key role in 5G-Advanced~\cite{RIS_standardization} as well as in 6G wireless networks~\cite{Samsung}. 
 Another application area where the usage of \acp{ris} is the focus of considerable research attention is \ac{rf} localization~\cite{bjornson2021reconfigurable}. The conventional task in such systems considers a mobile device which determines its position based on the impinging signals received from several terminals whose locations are known. The commonly studied capability of \acp{ris} in facilitating RF localization follows from its reliance on forming multiple signals received from known locations. \acp{ris} can thus create additional signal propagation paths for their impinging signals without increasing the number of transmitting terminals, as well as overcome \ac{nlos} conditions in an energy-efficient manner~\cite{Keykhosravi2022infeasible,elzanaty2021reconfigurable}. The growing popularity of \ac{ris}-enabled/-aided \ac{rf} localization is also attributed to the fact that this application area is naturally associated with wireless communications devices, where the \ac{ris} research is being recently highly established.

 This article focuses on another emerging technology: the \ac{isac} paradigm. This technology unifies wireless communications and \ac{rf} sensing, since both applications are associated with radiating \ac{em} waves and are expected to be simultaneously employed by a multitude of mobile devices \cite{ma2020joint}. As such, \ac{isac} is envisioned to be an integral part of future wireless systems, especially when operating at the millimeter-wave (mmWave) and terahertz (THz) frequency bands \cite{Liu_ISAC,liu2021integrated,Zhang_ISAC,Cui_ISAC}. However, signal propagation at these high frequencies is quite challenging due to the penetrating pathloss, which can be so severe that the NLoS paths may be too weak to be of practical use, preventing reliable communication and, in certain cases, sensing. 
\ac{isac} systems share many similar aspects, in both operation and challenges, with the more established \ac{ris} domains of wireless communications and RF localization. In particular, data communication is one of the two functionalities incorporated in \ac{isac} systems. Yet, while \ac{rf} localization can be viewed as a form of sensing, the way the propagation of \ac{em} waves is used typically differs from that in \ac{isac} systems. In particular, ISAC commonly focuses on a wireless receiver (or transceiver) utilizing impinging \ac{rf} signals to sense an unknown environment, similar to what happens in radar systems. Nonetheless, the ability of \acp{ris} to generate controllable desired reflection patterns was recently shown to facilitate sensing applications, e.g.,~\cite{buzzi2021foundations}. This indicates that the combination of \acp{ris} with \ac{isac} systems may contribute to the development of future wireless networks and their mobile devices. 

In this article, we review the applications of \acp{ris} for \ac{isac} systems, as well as the signal processing challenges and the offered gains which arise from the integration of these two emerging technologies. We present up-to-date research directions of such synergies, contribute new fundamental insights, and highlight exciting research opportunities arising from the usage of \acp{ris}. For this goal, we begin by reviewing the fundamentals of the \ac{ris} technology which are relevant for \ac{isac} applications. We briefly describe the \ac{ris} hardware architectures, focusing mostly on conventional passive architectures~\cite{risTUTORIAL2020,WavePropTCCN}, as well as recent hybrid passive-active technologies \cite{alexandropoulos2021hybrid}. We then discuss the key capabilities of \acp{ris} in shaping wireless environments, and their associated use cases, based on which we establish a generic model for the signals involved in \ac{ris}-aided \ac{isac} systems that we specialize in the subsequent sections. We continue by overviewing the existing methodologies for utilizing \acp{ris} for sensing applications, focusing specifically on radar sensing, which is the closest functionality to \ac{isac} systems. While the wireless communications functionality, e.g., rate or \ac{sinr} improvement, considered in \ac{isac} is similar to that commonly treated in the \ac{ris} literature, we provide new perspectives on the sensing aspect of \acs{ris}-aided \ac{isac}, which is typically geared towards radar-type operations, e.g., detecting targets and probing an unknown environment. We discuss the different \ac{ris}-aided sensing configurations such as monostatic active sensing, where the same device that radiates the signal is the one probing the environment, as well as bistatic sensing, where a passive sensing receiver processes the impinging signals to estimate parameters for targets in the environment. We discuss transmit beamforming to show the ability of \acs{ris} to enhance the target illumination power and improve sensing performance. The gains offered by \acp{ris} for sensing in harsh and \ac{nlos} environments are showcased via intuitive numerical examples.

The main bulk of the articles considers \ac{ris}-empowered \ac{isac}. We discuss for the first time the dependence of \ac{isac} on the {\it coupling} between the communication and sensing channels, proving that the more coupled the channels are, the more gains one can achieve via joint design of communications and sensing. We then show that, in the context of \ac{isac}, \acp{ris} provide a means to control the level of coupling between the channels. This reveals a non-trivial core benefit of combining \acp{ris} with \ac{isac}, in which signal processing techniques play a profound role. We accompany our discussion with a beamforming design example in multi-antenna \ac{isac} systems with a single-antenna communication receiver. For this example, we draw observations on the role of coupling in the \ac{isac} performance. We then show how the integration of \acp{ris} affects the level of coupling, and in turn, the performance gains of a joint \ac{snc} design. Based on this fundamental observation, we present a novel design for simultaneously providing high-rate communications, while achieving accurate estimation of remote targets, which leverages the capabilities of \ac{ris}-empowered \ac{isac}. We also discuss an \acs{ris}-aided \acs{isac} design that guarantees a prescribed \ac{sinr} to a communication user and yields sensing waveforms with good cross-correlation properties, thus adequate for sensing and resolving targets. The article is concluded with a discussion on open challenges and future research directions associated with \ac{ris}-empowered \ac{isac} systems. Exploring these directions is expected to pave the way to a successful fusion of the concepts of smart wireless environments and \ac{isac}, further bringing forth the gains of joint holistic designs of \acp{ris} and \ac{isac}.

\section{Fundamentals of \acp{ris}} \label{sec:fund_ris}
\label{sec:basics}
In this section, we include the key features of the \acp{ris} technology, setting the ground for the description of their applications in the following sections in the context of \ac{isac}. We begin with a brief description of the \acp{ris} hardware architectures. Then, we elaborate on the requirements, capabilities, and use cases of \acp{ris} for controlling wireless propagation environments, based on which we formulate a generic signal model that is specialized in the rest of upcoming sections. 
%
%
\subsection{Hardware Architectures}
\label{ssec:Architectures}
\begin{figure*}[!t]
\centering
\subfigure[Simultaneous reflecting and sensing \ac{ris} architecture.]{ 
\label{fig:HRIS1} 
\includegraphics[width=0.35\linewidth, height=3.3cm]{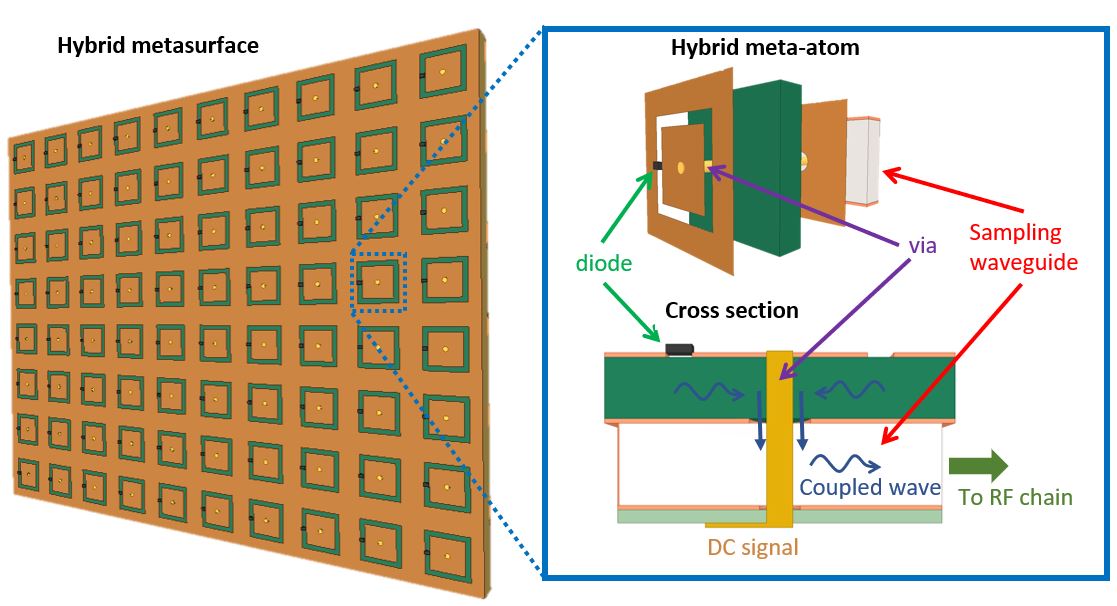}}
\quad
\subfigure[Reflected beampattern.]{ 
\label{fig:HRIS2} 
\includegraphics[width=0.3\linewidth, height=3.3cm]{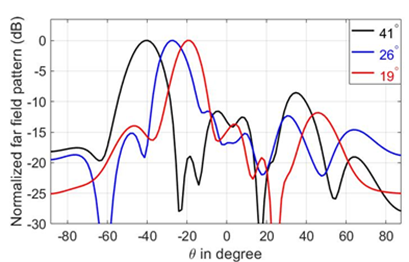}}
\quad
\subfigure[AoA detection.]{ 
\label{fig:HRIS3} 
\includegraphics[width=0.25\linewidth, height=3.3cm]{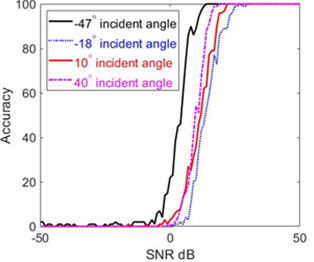}}
\caption{\label{fig:HRIS} The schematic of a hybrid passive and active \ac{ris} and full-wave simulation results adopted from \cite{alexandropoulos2021hybrid}. The results in $(b)$ and $(c)$ were obtained with an one-dimensional hybrid \ac{ris} having $24$ meta-atoms, and demonstrate the metasurface's capability to reflect towards desired angles, while accurately estimating the \acp{aoa} of its impinging signal. }
 \vspace{-0.3cm}
\end{figure*}

 \begin{tcolorbox}[float*=t,
    width=\columnwidth,
	toprule = 0mm,
	bottomrule = 0mm,
	leftrule = 0mm,
	rightrule = 0mm,
	arc = 0mm,
	colframe = myblue,
	colback = mypurple,
	fonttitle = \sffamily\bfseries\large,
	title = Passive versus Active and Hybrid RISs]	
	\label{Box:PassivevsActive}
By definition, \acp{ris} are metasurfaces with reconfigurable electromagnetic properties. The common usage of this terminology, which is also the one adopted in this article, refers to almost passive devices used as a sort of a full-duplex relay that can generate controllable scattering and reflection patterns without being capable of amplifying its impinging signals. An alternative usage of metasurfaces with controllable electromagnetic characteristics is as reconfigurable massive MIMO antennas, including the concept of dynamic metasurface antennas \cite{shlezinger2020dynamic}. The latter usage, which serves as a candidate technology for {\it holographic \ac{mimo}} \cite{huang2020holographic}, 
is intended as being a part of an active transceiver with dedicated RF chains and signal amplifiers (power and low-noise amplifiers for transmitters and receiver, respectively), and thus its capabilities and tasks are fundamentally different from that of passive \acp{ris}. 

The passive nature of \acp{ris}, which is used for programming wireless environments, gives rise to multiple design challenges, and the resulting metasurfaces are dependent on at least one distinct receiver device for sensing the environment and handling their configurations. To tackle these challenges, a hybrid passive and active \ac{ris} architecture have been proposed \cite{alexandropoulos2021hybrid}, consisting of response-reconfigurable meta-atoms placed on top of a waveguide that offers adjustable coupling of the impinging signals on them to one or more reception RF chains (see Fig.~\ref{fig:HRIS1}). The resulting architecture was shown to achieve desired reflection beampatterns (see Fig.~\ref{fig:HRIS2}), while providing sensing capability to locally estimate the \acp{aoa} at the same time (see Fig.~\ref{fig:HRIS3}). 
\end{tcolorbox}	
Broadly speaking, an \ac{ris} is a surface having multiple elements whose electromagnetic properties can be externally configured. We focus our description on passive \acp{ris}, which are comprised of passive reflecting elements that allow achieving controllable reflection patterns. Nonetheless, one may utilize \acp{ris} that are not purely passive, and possess some sensing capabilities on their own, as detailed in the box entitled \emph{Passive versus Active and Hybrid RISs} on the following page~\pageref{Box:PassivevsActive}.
 
 While \acp{ris} can be realized using conventional reflectarrays, a common implementation utilizes metasurfaces~\cite{CE_overview_2022}. Metasurfaces are two-dimensional arrays of artificial metamaterial elements, which can be based on, e.g., graphene or liquid crystals. The electromagnetic properties of these elements, referred to as meta-atoms, can be programmed, and their dimensions allow them to be densely packed, possibly within sub-wavelength spacing. The surface configuration is typically placed upon a dielectric substrate layer, whose role is to prevent energy leakage from impinging signals and their attenuation upon reflection. As a result, \acp{ris} may be fabricated in flexible shape and thickness, facilitating their deployment. In \ac{ris} architectures that do not act as purely reflectors, such as \acp{ris} that also sense the impinging signals \cite{alexandropoulos2021hybrid} (see the box on the top of this page) or refract portions of it \cite{mu2021simultaneously}, the substrate layer is replaced with a different substance, e.g., a waveguide exhibiting limited coupling to the elements~\cite{Alamzadeh2021ris}. 
 
 The ability to modify the signal propagation profile via deploying \acp{ris} is achieved using dedicated control circuitry for their dynamic configuration. The control circuitry can adapt the electromagnetic behaviour of each meta-atom separately, or of groups of meta-atoms, either in a binary (or discrete) fashion by incorporating one or several positive-intrinsic-negative (PIN) diodes in each element, or even in a continuous manner using, e.g., varactors. While the control circuit inevitably consumes power, the almost passive nature of the elements themselves (in the typical order of $\mu$Watts \cite{WavePropTCCN}) allows the overall architecture to be nearly-passive, typically consuming much less power than that needed by active transmitters/receivers/relays~\cite{pan2021reconfigurable}.
 
%
\subsection{Capabilities and Use Cases}
\label{ssec:Requirements}
\subsubsection*{Capabilities}
The \ac{ris} technology enables to reflect impinging EM waves in a controllable manner. In that sense, \acp{ris}, which are typically considered as means to control the wireless environment, can also be viewed as a form of passive relays \cite{huang2019reconfigurable}. In fact, they have several core advantages compared with conventional relays that utilize active transmitters. First, \acp{ris} naturally operate in a full-duplex fashion, an operation which increases the spectral efficiency, and may be challenging to achieve using conventional relays. Furthermore, the flexible planar shape of \acp{ris} facilitates their deployment, particularly in urban settings, e.g., on the facades of buildings \cite{di2019smart}. Finally, the limited control circuitry of \acp{ris}, which is based on the fact that they do not process the impinging signals and do not require RF chains, combined with their relatively simple fabrication, indicate that metasurfaces are likely to have lower cost compared with active relays \cite{pan2021reconfigurable}. 

While the passive nature of \acp{ris} leads to operational gains in spectral efficiency and cost, it also imposes some limitations on their capabilities. The fact that \acp{ris} do not sense the impinging signals indicate that they need to rely on an external device to configure their reflection pattern. This can be a \ac{bs} in a wireless communications setting, or some kind of a dedicated controller. The external device should have knowledge on the environment and the location of the transmitting entities, and can use this knowledge to control the wireless propagation by configuring the reflection patterns of possibly multiple \acp{ris}. \acp{ris} should be deployed with a dedicated control link and thus cannot be independent, relying on sophisticated optimization and signal processing carried out in real time from the RIS controlling device. 

Furthermore, the ability to shape the environment via \acp{ris} is typically modeled by representing each element as a controllable phase shifter. This relatively simplified model, which is widely adopted in the literature and also used in some of the subsequent sections in this article for simplicity, may not faithfully reflect their operation. First, the phase profile of each element is typically coupled with both its attenuation as well as the incident angle of the impinging signal \cite{chen2020angle}. Additionally, even when one can achieve a desired phase shift for a narrowband signal, this behaviour does not necessarily holds for wideband signals, since there is typically coupling and structured frequency selectivity at each element \cite{katsanos2022wideband}. Finally, in the presence of multipath propagation, where multiple reflections of the same signal are received from different incident angles, the ability of \acp{ris} to shape the propagation environment becomes notably more complex compared with the simplified phase-shifting model \cite{faqiri2022physfad}. Despite these inherent mismatches in the conventional phase shifter model of \acp{ris}, it facilitates signal processing designs and provides meaningful insights to the potential gains and usefulness of \ac{ris}-aided/-enabled wireless communications and sensing systems.

\subsubsection*{Use Cases} 
To date, \acp{ris} are mostly studied in the context of data communication via their reflective beamforming capability, and are considered to be a key enabling technology for future generations of wireless communications \cite{Samsung}. To this end, metasurfaces are expected to pave the way to the smart radio environment vision \cite{di2019smart}, bringing forth the ability to improve signal coverage and overcome harsh \ac{nlos} conditions, particularly in urban settings~\cite{di2019smart}; boost the energy efficiency of the wireless network~\cite{huang2019reconfigurable}; safeguard legitimate links against eavesdropping \cite{georgeRISpls2021}; ameliorate the throughput, fairness, and reliability of wireless communications~\cite{RISE6G_COMMAG}; and facilitate interference mitigation in, possibly dense, multi-user communication systems~\cite{huang2019reconfigurable,wu2019beamforming}.   

\acp{ris} are leveraged to form desirable reflection patterns as means of conveying information in wireless communication systems. This is achieved by embedding additional information bits in the reflection pattern of the \ac{ris} as a form of {\it index modulation} \cite{basar2020reconfigurable}. Alternatively, one can utilize a transmitter (Tx) that radiates a simple waveform with minimal processing at the \ac{ris}, which reflects in a manner that depends on the message. The latter technique, coined {\it reflection modulation} \cite{reflection_pattern_modulation2021}, is considered to be quite promising especially for high-frequency communications, where designing cost and power efficient Txs utilizing conventional digitally modulated waveforms is challenging. 

While the main focus of the different use cases of \acp{ris} is geared towards wireless communications, their ability to shape the wireless propagation environment, in a manner which is sustainable and energy efficient, may contribute to other applications utilizing EM radiation. One such application is wireless power transfer~\cite{yang2021reconfigurable}, where \acp{ris} can be used to focus the radiated energy towards the desired users, while mitigating radiation in undesired directions (i.e., energy pollution). 


A third family of \ac{ris} use cases involves the usage of EM waveforms for positioning and sensing tasks. The most commonly studied usage of \acp{ris} for sensing considers \ac{rf} localization \cite{george_henk_ICC2021,Keykhosravi2022infeasible,elzanaty2021reconfigurable}. In this context, a mobile device aims at localizing its position based on signals transmitted from terminals whose location are known. When the device receives sufficient signals and is within the \ac{los} of the known terminals, it can identify its position based on geometric considerations. The incorporation of \acp{ris} for RF localization thus brings forth the ability to generate additional reflections which can be utilized for improving localization performance. Consequently, \ac{ris}-aided/-enabled localization can improve accuracy as well as facilitate operation in settings with a limited number of transmitting terminals and \ac{nlos} environments. In fact, \acp{ris} can enable the localization of users even in cases where there is no access point or BS available in the system \cite{Keykhosravi2022infeasible}. Additional sensing-related applications of \acp{ris} which have been recently explored in the literature include radio/environmental mapping \cite{Kimmo_Popovski_2021}, RF sensing~\cite{hu2020reconfigurable}, and radar~\cite{buzzi2021foundations}.

The fact that \acp{ris} can be used for both sensing and communications gives rise to use cases related to applications involving simultaneous sensing and communications, either as means for facilitating the functionalities' co-existence for spectrum sharing applications~\cite{he2022risassisted} as well as for dual-function systems, i.e., \ac{isac}~\cite{sankar2021joint,sankar2022beamforming,sankar2022beamforming2}. The family of \acp{ris}-empowered ISAC applications is the focus of this article and will discussed extensively in the subsequent sections. 

\subsection{Signal Modeling}
\label{ssec:Model_Generic}
To gradually reveal the potential benefits of incorporating \acp{ris} in \ac{isac} systems, we formulate a generic yet simplified signal model, which is repeatedly used in the relevant literature and specialized in the sequel. We consider the transmission of a (possibly) dual-function signal from a transmit array consisting of $\NTx$ antenna elements. The signal, denoted by $\SigTx(t)\in \mathbb{C}^{\NTx}$ with $t$ being the time index, can encapsulate both a radar waveform as well as a digital communication message. Such dual-function signaling (or coordinated signaling) can be achieved via the usage of communication signals for probing, or the embedding of messages in a radar waveform (see, e.g., \cite{ma2020joint} for an indicative scheme).

In \ac{ris}-empowered wireless systems, the signal propagates through a wireless channel whose response can be manipulated via the configurations of the RISs. Let us consider an \ac{ris} with $\NRis$ elements, parameterized by the vector $\myVec{\phi}\in\mathbb{C}^{\NRis}$ (including the reconfigurable responses of its elements), whose reflections are received by a radar receiver (Rx) with $\NRxRad$ antenna elements and by a communications Rx with $\NRxCom$ antenna elements. The radar Rx may be co-located with the Tx, as in monostatic radar and \ac{dfrc} systems \cite{chiriyath2017radar_commun}, or be a separate entity, as in bistatic radar and other passive sensing applications. The received radar signal  $\SigRxRad(t)\in \mathbb{C}^{\NRxRad}$ and the received communications signal $\SigRxCom(t)\in \mathbb{C}^{\NRxCom}$ can be respectively expressed as
\begin{equation}
    \label{eqn:Rx}
    \SigRxRad(t) = \ChRad(\SigTx(t); \myVec{\phi}); \qquad \SigRxCom(t) = \ChCom(\SigTx(t); \myVec{\phi}).
\end{equation}
where $ \ChRad(\cdot ; \myVec{\phi})$ and $ \ChCom(\cdot ; \myVec{\phi})$ are the (stochastic) sensing and communications channel, respectively, which are both parameterized by the \ac{ris}. For instance, using the simplified narrowband cascaded channel model of phase-shifting \acp{ris} without a direct link \cite{huang2019reconfigurable}, $ \ChCom(\cdot ; \cdot)$ can be approximated as
\begin{equation}
     \ChCom(\SigTx(t); \myVec{\phi}) = \myMat{H}_{\rm RIS-Rx}{\boldsymbol \Phi} \myMat{H}_{\rm Tx-RIS} \SigTx(t) + \myVec{z}(t),
     \label{eqn:Cascaded}
\end{equation}
where $\myMat{H}_{\rm Tx-RIS}$ and $\myMat{H}_{\rm RIS-Rx}$ represent the Tx-\ac{ris} and \ac{ris}-Rx channels, respectively, ${\boldsymbol \Phi} = {\rm diag}(\myVec{\phi}) \in \mathbb{C}^N$ is the diagonal matrix containing the tunable phase shifts of the \ac{ris}, and $\myVec{z}(t)$ is additive white Gaussian noise modeling the reception thermal noise. The elements of $\myVec{\phi}$ are usually modeled as unit modulus, i.e., vary solely in phase. Note that more realistic models capturing the intertwinement between each meta-atoms tunable phase and amplitude are available \cite{Abeywickrama_2020}.   

We next utilize the 
 model \eqref{eqn:Cascaded} to discuss signal processing techniques for applying \acp{ris} in sensing as well as in \ac{isac} systems. Our motivation for adopting this simplified representation stems from its ability to facilitate the presentation of the system operation, allowing us to draw insights and provide indications on the benefits of \ac{ris}-aided \ac{isac}.

\section{Sensing with \acp{ris}}
\label{sec:Sensing}


Sensing is a fundamental task in radar, imaging, and wireless communications. Sensing problems in radar and wireless communications boil down to estimating parameters of targets or point scatterers from received signals (e.g., angle, range, and Doppler shift). Sensing of targets or point scatterers is challenging when operating at high frequencies (e.g., in mmWave bands) due to the propagation loss and the usually absence of the direct visibility of them. In such harsh propagation environments, \acp{ris} may be leveraged to introduce virtual paths and/or intentional geometric delays. 
%
%

\begin{figure*}
    \centering
    \includegraphics[width=\columnwidth]{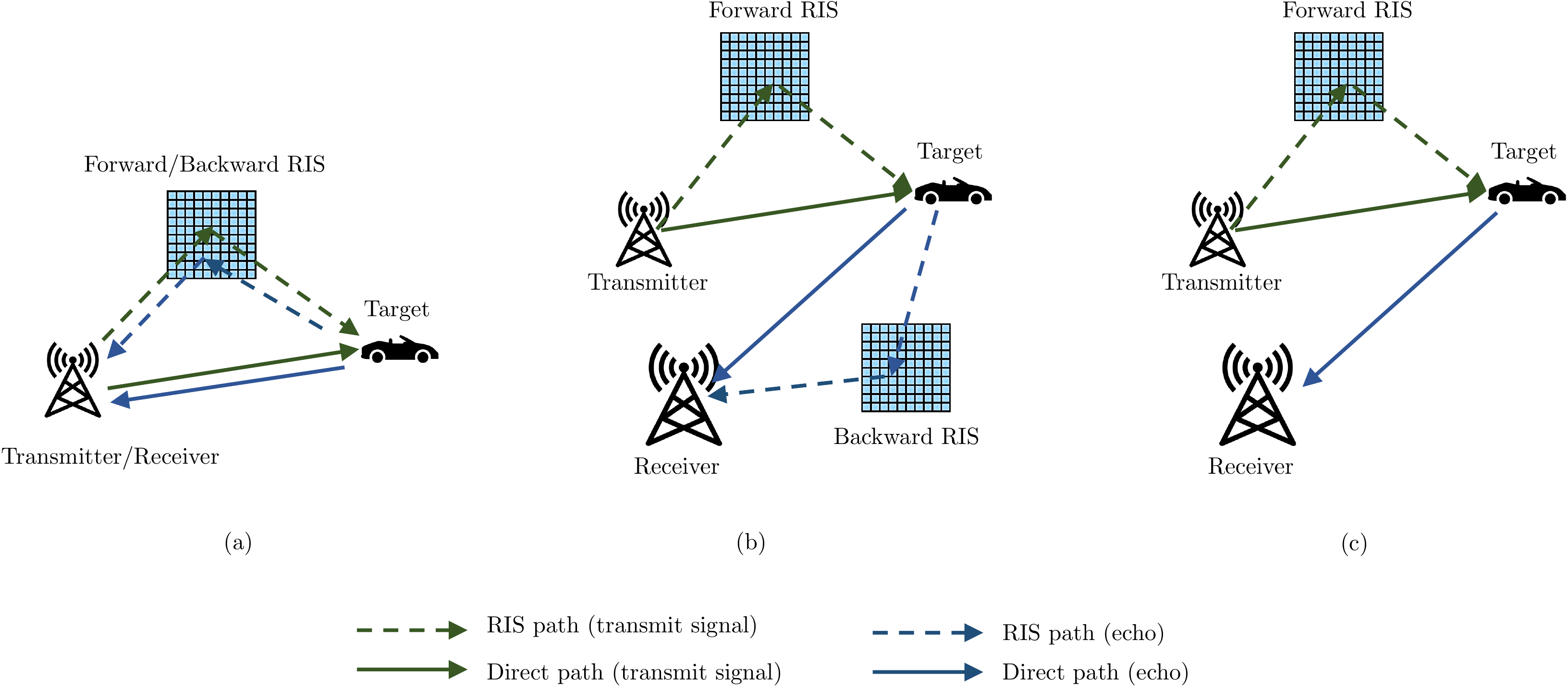}
    \caption{RIS-aided sensing configurations: (a) Monostatic; (b) Bistatic with \acp{ris} handling either forward or backward/echo relaying. (c) Bistatic with one \ac{ris} assigned to handle only forward relaying.}
    \label{fig:ris_aided_sensing}
\end{figure*}

\begin{tcolorbox}[float*=t,
    width=\columnwidth,
	toprule = 0mm,
	bottomrule = 0mm,
	leftrule = 0mm,
	rightrule = 0mm,
	arc = 0mm,
	colframe = myblue,
	colback = mypurple,
	fonttitle = \sffamily\bfseries\large,
	title = \acs{ris}-Aided Sensing Configurations]	
	\label{Box:ris_sensing}
The two common configurations for sensing are the \emph{monostatic} and \emph{bistatic} configurations. In a monostatic sensing system, the transmitter and receiver are collocated. In a bistatic sensing system, the transmitter and receiver are placed at separate locations. RIS-aided sensing systems may have both \emph{Forward \acs{ris}} and \emph{Backward \acs{ris}} to create additional propagation paths for the transmitted signals and their echoes after bouncing on targets/obstacles.
Figure~\ref{fig:ris_aided_sensing}(a) illustrates a monostatic sensing system with a single \acs{ris}. In the absence of the direct path, the transmit signals may illuminate the target via the \acs{ris}, and the echoes may reach the receiver again via the \acs{ris}. In Fig.~\ref{fig:ris_aided_sensing}(b), we illustrate an \acs{ris}-aided bistatic sensing scenario, where one forward \acs{ris} and one backward \ac{ris} assist the sensing mechanism. The target is illuminated by both the direct path as well as the additional paths via the \acs{ris}, i.e., the link from the transmitter to the target via the Forward \acs{ris}. Naturally, the paths due to \acs{ris} will be weaker as they undergo double pathloss, however, those will be the only enablers for sensing in the absence of the direct path. In addition, for certain \acs{ris} placements (e.g., when the Forward \acs{ris} is close to the target or the transmitter), it could be that the direct path and \acs{ris}-based one are of comparable strengths, leading to a better target illumination power. At the receiver side, we may experience the same problem with the targets being directly invisible to the receiver. In such scenarios, a Backward \acs{ris} can further ensure that the echo from the target reaches the receiver. Note that, when only relying on \acp{ris} without direct paths, the signal undergoes an inevitable quadruple pathloss. An important design problem in such \acs{ris}-aided sensing systems is to design the transmit beamformer to ensure that sufficient power reaches the targets. Figure~\ref{fig:ris_aided_sensing}(c) shows a simplified version of the bistatic sensing system in Fig.~\ref{fig:ris_aided_sensing}(b) without Backward \acs{ris}.
\end{tcolorbox}

Traditionally, sensing is performed by transmitting a probing signal to illuminate a target. The reflected echoes from the targets are then processed at the radar receiver. It is important that sufficient power reaches the targets so that these are well illuminated to begin with, while power of the reflected echoes depends on the target gain (also referred to as the \textit{radar cross section}. When there is no direct path between the target and the Tx, then hardly any power reaches the target. In such scenarios including targets in \ac{nlos}, \acp{ris} can be highly useful. \ac{ris}-aided sensing corresponds to several different configurations, as detailed in the box entitled {\em \ac{ris}-Aided Sensing Configurations} on the next page~\pageref{Box:ris_sensing}. For ease of presentation in what follows, we consider the case of a single point target to discuss the benefits of deploying \acp{ris} for its sensing. We particularly focus on a bistatic sensing system having only a Forward \acs{ris}, as illustrated in Fig.~\ref{fig:ris_aided_sensing}(c). In the following, we overview \ac{ris}-aided sensing under such settings by first presenting the received signal model. Then, we describe the active and passive transmit beamforming mechanisms for RIS-based sensing, identifying the regimes in which \acp{ris} are most beneficial. We also consider the problem of target and parameter detection, where we showcase the ability of \acp{ris} to generate phase profiles with improved target power, which is most notable in harsh and \ac{nlos} settings, which directly translates into improved target detection capability.

\subsection{Sensing Signal Model}
We commence with the specialization of the signal model in \eqref{eqn:Rx} to the focused bistatic \ac{ris}-aided sensing setup in Fig.~\ref{fig:ris_aided_sensing}(c). We consider Tx and Rx arrays with $\NTx$ and $\NRxRad$ elements and an \acs{ris} array with $N$ elements. Let us denote the angular position of the point target in the far field with respect to the transmit array and the \ac{ris} array by $\theta_1$ and $\theta_2$, respectively. In addition, let $s\left(t\right) \in \mathbb{C}$ denote the discrete-time baseband sensing waveform at time $t$. We assume that the sensing waveform has unit power, and that it is precoded using a beamformer $\bw \in \mathbb{C}^{\NTx}$. Then, the transmit signal is $\SigTx \left(t\right) = \bw s\left(t\right) \in \mathbb{C}^{\NTx}$. 

The direct Tx-target and Rx-target channels are modeled as \ac{los} channels and are given by ${\alpha _t}{{\myVec{a}}_t}\left( \theta_1  \right)$ and ${\alpha _r}{{\myVec{a}}_r}\left( \theta_1  \right)$, where $\alpha_t$ and $\alpha_r$ are the complex path gains, and ${{\myVec{a}}_t}\left( \theta_1  \right) \in \mathbb{C}^{\NTx}$ and ${{\myVec{a}}_t}\left( \theta_1  \right) \in \mathbb{C}^{\NRxRad}$ are the transmit and receive array steering vectors towards the target directions~$\theta_1$ with ${{\myVec{a}}_t^H\left( \theta_1  \right)}{{\myVec{a}}_t\left( \theta_1  \right)}= \NTx$ and ${{\myVec{a}}_r^H\left( \theta_1  \right)}{{\myVec{a}}_r\left( \theta_1  \right)}= \NRxRad$. Let ${\boldsymbol b}(\theta_2)$ denote the steering vector of the \ac{ris} towards the target at direction $\theta_2$. Then, the overall Tx-target and Rx-target channels are expressed as:
\begin{equation}\label{eq:RIS_channels}
\begin{gathered}
  {{\myVec{h}}_t} = \underbrace{{\alpha _t}{{\myVec{a}}_t}\left( \theta_1  \right)}_{\text{direct path}} + \underbrace{{{\mathbf{G}}_t}{\mathbf{\Phi b}}\left( \theta_2  \right)}_{\text{RIS path}}, \hfill \\
    {{\myVec{h}}_r} = \underbrace{{\alpha _r}{{\myVec{a}}_r}\left( \theta_1  \right)}_{\text{direct path}} + \underbrace{{{\mathbf{G}}_r}{\mathbf{\Phi b}}\left( \theta_2  \right)}_{\text{RIS path}}, \hfill \\
\end{gathered}
\end{equation}
where $\mG_t =  \beta_t{{\myVec{a}}_t}(\omega_t) {\boldsymbol b}^H(\omega_t) \in \mathbb{C}^{\NTx \times N}$ and $\mG_r =  \beta_r{{\myVec{a}}_r}(\omega_t) {\boldsymbol b}^H(\omega_t) \in \mathbb{C}^{\NRxRad \times N}$ are the channels between the \ac{ris} and Tx and between the \ac{ris} and Rx with $\omega_t$ being the direction of the path between the Tx and the \ac{ris}. Here, $\beta_t$ and $\beta_r$ are the overall attenuation of the \ac{ris} paths. While $\alpha_t$ is inversely proportional to the squared distance between the Tx and the target, $\beta_t$ is inversely proportional to the product of the squared distance between the Tx and \acs{ris} and the \ac{ris} and the target. Hence, $\beta_t$ would be, in general, much smaller than $\alpha_t$.
The received echo signal can be thus represented as
\begin{equation}\label{eq:echo_model}
{\myVec{y}_s}\left( t \right) = {{\myVec{h}}_r}{\myVec{h}}_t^H{\vw}s\left( t \right) + {{\myVec{z}}_s}\left( t \right),
\end{equation}
where ${{\myVec{z}}_s}\left( t \right)$ represents the zero-mean additive white Gaussian noise with variance~$\sigma_s^2$. Note that the overall received echo signal comes from the two directions, i.e., from the target reflection angle $\theta_1$ and the RIS reflection angle $\omega_t$. There are four propagation paths in the sensing channel, namely, the BS-target-BS, BS-RIS-target-RIS-BS, BS-RIS-target-BS, and the BS-target-RIS-BS channels. Compared to conventional downlink communication scenarios, the three newly added paths provide not only extra channel gains that can be used for sensing, but also offer an additional dimension to measure the target. Specifically, by estimating the key parameters contained in the four paths, i.e., $\theta_1$ and $\theta_2$, one can \textit{localize} the target. It is noted that the effect of the echoes from the \ac{ris} on the Rx array will be relatively small, hence, we hereinafter assume that their contribution is negligible.
\subsection{Transmit Beamforming}
The target illumination power, which is defined as the power of the signal at the target due to the transmit sensing waveform $\SigTx \left(t\right)$, can be mathematically expressed as follows:
\begin{equation*}
 {\rm power}(\vw, \boldsymbol{\Phi}) \triangleq \mathbb{E}[|\vh_t^H\SigTx \left(t\right)|^2] = \vh_t^H \vw\vw^H \vh_t.   
\end{equation*}
For a given target direction, the target illumination power not only depends on the precoder $\vw$, but also on the reflection pattern of the \acs{ris}. By optimally designing the precoder and the \acs{ris} reflection pattern, one can improve the effective target illumination power, which in turn improves sensing performance of the system. Thus, we can maximize the target illumination power with respect to $(\vw, \boldsymbol{\Phi})$:
\begin{align}
    &\underset{\vw, \boldsymbol{\Phi}}{\text{maximize}} \,\,\,\, {\rm power}(\vw, \boldsymbol{\Phi}) \nonumber\\
    &\text{s.t.} \quad \quad\; \|\vw\|^2_2 \leq 1 \nonumber\\
    &\quad \quad\quad\quad |\phi_i| = 1, \, i=1,\ldots,N, \label{eq:power_opt}
\end{align}
where the constraint on $\vw$ is due the transmit power budget, which we set to $1$ without loss of generality. The unit modulus constraint on $\phi_i$ is based on the simplified phase-shifting model of \acp{ris}, as discussed in Section~\ref{ssec:Model_Generic}. This optimization problem is formulated for a single target for simplicity, and can be easily extended to multiple targets with the objective of maximizing the worst-case illuminated power over all of them. As observed, this design problem is non-convex in the variables $(\vw, \boldsymbol{\Phi})$, but it can be sub-optimally solved using alternating optimization: by fixing one variable and solving with respect to the other and vice versa, and iterating this procedure till convergence. 

Without the \ac{ris} and with only the direct path to/form the target, the channels in \eqref{eq:RIS_channels} will not have the second terms. This of course implies that there will be no \ac{ris} phase profiles to be optimized. In this case, the target illumination power can be trivially maximized by aligning $\vw$ to the channel as
\begin{align}
\vw^\star = \frac{\myVec{a}_t(\theta_1)}{\|\myVec{a}_t(\theta_1)\|_2},
\label{eq:precoder}
\end{align}
which is the well-known matched-filtering beamformer.

In order to understand when an \acs{ris} is actually useful for improving the target illumination power and when it is not, we next discuss the beamforming gain offered by the \ac{ris} in cases with and without the direct path.

\textbf{RIS beamforming gain with no direct path:} Whenever there is no direct path, \acp{ris} can play an important role in illuminating the target, when designed to focus the energy from the Tx to it. To do so, the sensing beamformer is aligned to the channel between the Tx and \ac{ris}, i.e., $\vw = \frac{\va_t(\omega_t)}{\|\va_t(\omega_t)\|_2}$. This design implies that all Tx energy is beamformed towards the \acs{ris}. In this case, the useful signal reaching the target is:
\begin{align}
\vh_t^H\SigTx\left(t\right) \propto {\boldsymbol b}^H(\theta_2) \bar{\boldsymbol{\Phi}} {\boldsymbol b}(\omega_t).    
\end{align}
Hence, the optimal choice of $\boldsymbol{\Phi}$ to maximize the target illumination power is to choose $\phi_i$ for $i=1,\ldots,N$ as follows:
\begin{align}
    \bar{\phi}_i = \exp\{-j \, \text{angle}([{\boldsymbol b}^H(\theta_2) \odot {\boldsymbol b}(\omega_t)]_i)\}.
    \label{eq:phase_shift}
\end{align}
where symbol $\odot$ is the Hadamard product. The total illumination power is then $\mathbb{E}[|\vh_t^H\SigTx\left(t\right) |^2] = \sigma_\beta^2\NTx N^2$, where $\sigma_\beta^2 = \mathbb{E}[|\beta_t|^2]$ is the average strength of the \ac{ris} path. Without the direct path, we can see that, by using an \acs{ris}, a beamforming gain of $N^2$ can be achieved due to the RIS, but only an array gain of $\NTx$ due to the Tx array.

\textbf{RIS beamforming gain with the direct path:} Suppose we choose an orthogonal precoding matrix $\mW$ to transmit probing signals as $\SigTx(t) = \mW \vs(t) $ so that $\mW\mW^H = {\mI}$ and $\mathbb{E}\{\vs(t)\vs(t)^H\} = \mI$. In other words, suppose we transmit isotropically and choose the \ac{ris} reflection pattern as in \eqref{eq:phase_shift}. Then, the target illumination power simplifies as follows:
\begin{align}
    \mathbb{E}[|\vh_t^H\SigTx(t)|^2] &= \mathbb{E}[\vh_t^H\vh_t] \nonumber \\
    &= \sigma_\alpha^2 \NTx + \sigma_\beta^2 \NTx N^2,
\end{align}
where $\sigma_\alpha^2 = \mathbb{E}[|\alpha_t|^2]$ is the strength of the direct path. 
Normally, the strength of the \ac{ris} path is typically weaker than the strength of the direct one. For instance, let $\sigma_\beta^2 = \rho \sigma_\alpha^2$ with $\rho < 1$, i.e., the \ac{ris} path is $\rho$ times weaker than the direct path. When choosing an \ac{ris} with $N > \frac{1}{\sqrt{\rho}}$, the RIS path will be strengthened more than the direct one. This is due to the beamforming gain of $N^2$ offered by the \ac{ris}. Although difficult to quantify, the gain from the \ac{ris} deployment and the target illumination power will be usually higher when the Tx beamformer and \ac{ris} phase shifts are optimally chosen, i.e., by finding the optimum of the joint design problem in~\eqref{eq:power_opt}.

\subsection{Target Detection and Parameter Estimation}\label{sed:det}
In this subsection, we show the ability of \acp{ris} to enhance target illumination power, which is of particular interest in \ac{nlos} conditions, as discussed above, since it directly translates into improved sensing capability. We focus on the Rx side of the \acs{ris}-aided sensing system in what follows. In particular, the signal at the target is reflected and received by the receive array. We can thus rewrite the received signal model as
\begin{align}
   {\myVec{y}_s}(t) &= \eta{{\myVec{a}}_r}(\theta_1) \vh_t^H\SigTx(t) + {{\myVec{z}}_s}\left( t \right),
    \label{eqn:active_sensing_single_target}
\end{align} 
where $\eta$ is the target gain (models the radar cross section as well as the path attenuation between the target and Rx) with variance $\sigma_\eta^2$. We assume that the Rx uses a matched filter to $ {{\myVec{a}}_r}(\theta_1)$. In this ideal case, the signal-to-noise ratio (SNR) at the output of the matched filter is given by
\[
{\rm SNR} = \frac{\NRxRad \sigma^2_\eta}{\sigma_s^2} {\rm power}(\vw, \boldsymbol{\Phi}),
\]
i.e., it depends on the target illumination power.  

We can detect the presence or absence of a target using the Neyman-Pearson detector with a generalized likelihood ratio test (GLRT). For RIS-aided sensing, GLRT with respect to the target amplitude to determine the presence of a target (hypothesis $H_1$) and its absence (hypothesis $H_0$) is: 
\[
\frac{|{\myVec{y}_s}^H{\myVec{a}}_r|^2}{\sigma_s^2} \underset{{H_0}}{\overset{H_{1}}{\gtrless}} \gamma,
\]
where $\gamma > 0 $ is the detection threshold, which is set in order to obtain the constant false alarm rate $P_f = e^{-\gamma}.$ For a $|\eta|^2$ that is non fluctuating, the detection probability is~\cite{buzzi2021foundations}:
\[
P_d = 
Q_1(\sqrt{2{\rm SNR}}, \sqrt{2\gamma}) 
\]
with $Q_1(\cdot,\cdot)$ being the Marcum $Q$-function. Since the received echoes do not depend on the \ac{ris}, the optimal Rx filter (in terms of SNR or $P_d$) is the matched filter. 

Standard direction finding methods~\cite{van2004optimum}, e.g., subspace-based methods or beamforming, can be used to find the target bearing angle. The \ac{crb} gives us a lower bound on the variance of an unbiased estimator for the target bearing angle. It also gives us a baseline on the performance of the direction estimator. Suppose now that the receive array is conjugate symmetric, i.e., ${\dot{\myVec{a}}}_r(\theta)^H{\myVec{a}}_r(\theta) = 0$ where ${\dot{\myVec{a}}}(\theta) = \partial{\myVec{a}}_r(\theta)/\partial\theta.$ Then, the \ac{crb} (conditioned on echo signal and the target gain) with respect to the angle $\theta$ is given by~\cite{van2004optimum}:
\begin{align} \label{eq:crb}
    {\rm CRB}(\theta_1) &= \frac{\NRxRad}{2T \|\dot{\myVec{a}}\|^{2}_2} \left[\frac{1}{\rm SNR} + \frac{1}{({\rm SNR})^2}\right]  \\
    &= \frac{1}{2T\|{\dot{\myVec{a}}}\|^{2}_2} \left[\frac{\sigma_s^2}{\sigma_\eta^2{\rm power}(\vw,{\boldsymbol \phi})} + \frac{\sigma_s^4}{\NRxRad\sigma_\eta^4{\rm power}^2(\vw,{\boldsymbol \phi})}\right], \nonumber
\end{align}
where $T$ denotes the number of samples/measurements of the received signal. This \ac{crb} expression reveals that increasing the illuminated power directly translates to improved target identification capability, i.e., leads to a lower \ac{crb}. Consequently, the fact that an \ac{ris} can enable target illumination in \ac{nlos} settings, as previously discussed, implies that it can notably facilitate sensing in such challenging scenarios. To achieve the gains of \ac{ris}-enabled sensing, one should design the \ac{ris} reflection pattern to maximize the illuminated power, e.g., following the derivation in \eqref{eq:phase_shift} for the considered exemplified setting. 

\begin{figure*}
    \centering
    \includegraphics[width=\columnwidth]{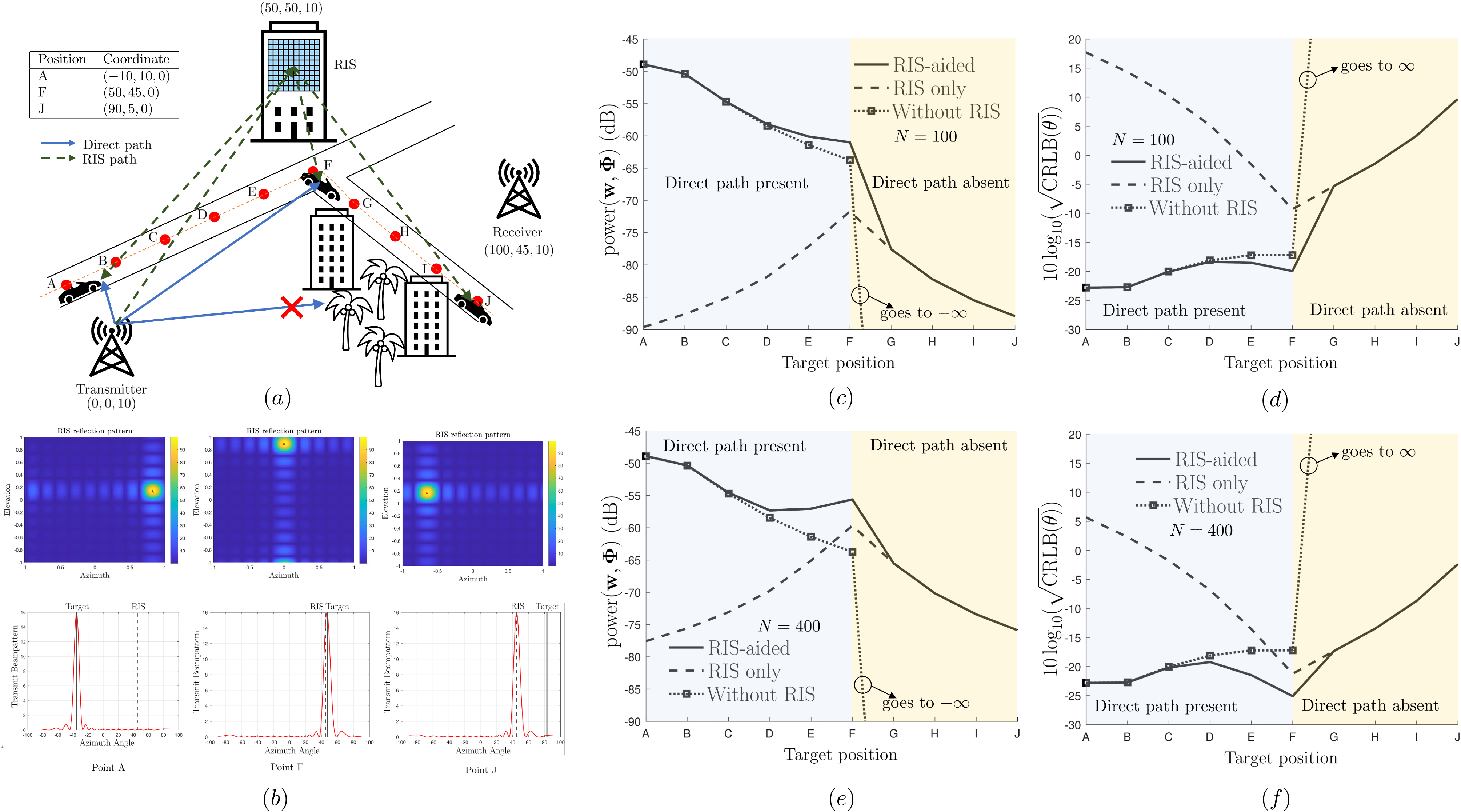}
       \caption{Transmit/reflection beamforming patterns and CRB for the considered RIS-aided sensing system in Section~\ref{sec:example}. (a) The point target (illustrated as a car) is moving on the road along the points A, B, C, up to point J. There is no direct path from points G till J. (b) Transmit beampatterns and RIS reflection pattern for different target locations. The symbol {\color{red}$\star$} indicates the target location with respect to the RIS phase profile. (c) and (e) Target illumination power for an \acs{ris} with $N=100$ and $400$ elements, respectively. (d) and (f) CRB with a $100$- and $400$-element \acs{ris}, respectively.}
    \label{fig:ris_aided_sensing2}
\end{figure*}

\subsection{Numerical Example}\label{sec:example}
We now illustrate the beamforming and sensing performances of an \acs{ris}-aided sensing system in scenarios with and without the direct Tx-target path. Specifically, we consider a setting in which the point target (illustrated as a car) is moving on the road along points A, B, C, up to the point, as illustrated in Fig.~\ref{fig:ris_aided_sensing2}(a).  The target is illuminated by both the direct path and the path via the RIS up to point F. For the points after F, the direct path is assumed to be completely blocked due to obstructions (such as trees or buildings). In essence, for the points G up to J, the target is illuminated only via the path introduced by the RIS. We jointly optimize the transmit beamformers at the Tx and the phase profiles at the RIS to achieve the maximum target illumination power according to \eqref{eq:power_opt}. Due to the coupling between the optimization variables (i.e., the precoder and the phase shifts), the resulting optimization problem is non-convex. We solve is in an alternating fashion by designing the precoders for a given phase shift and vice-versa~\cite{sankar2022beamforming}. In our conducted simulations, we have modeled each link as an LoS channel with a pathloss exponent of $2.5$ for the direct path and of $2.2$ for the \acs{ris} paths.

In Fig.~\ref{fig:ris_aided_sensing2}(b), we show the transmit beam pattern and the reflection pattern at the \acs{ris} when the target is at locations A, F, and J. We can observe that the \acs{ris} is always forming a reflective beam towards the target direction irrespective of whether the target is near or far. This is intuitive since the \acs{ris} attempts to focus whatever energy that it is receiving towards the target in order to increase the overall illumination power. It can be also seen that the transmit beamforming pattern is dependent on the target location. Specifically, at point A, the target is very close to the Tx. Hence, the transmit pattern has a single strong beam towards the target. In this setting, since the distance of the path of the Tx-\acs{ris}-target link is large, the gain offered by the \acs{ris} in improving the strength of the path via the RIS is not that significant. 
For Point F, the target is very close to the \acs{ris}. In this setting, it turns out that by a proper \acs{ris} phase profile configuration, it is indeed possible to obtain a sufficiently strong path via the metasurface, which is quite useful. Hence, the transmit beamformer is having prominent peaks towards both the \acs{ris} and the target. Finally, for point J, the direct path from the Tx to the target is blocked due to the presence of a building. The precoder thus concentrates all its energy towards the \acs{ris} to illuminate the target via the Tx-\acs{ris}-target link.

We now further analyze the benefits of the \acs{ris} deployment for sensing by evaluating the CRB of the estimator of $\theta_1$. As before, we assume that the target is moving along the path through points A till J. We consider three different scenarios for comparison: setting (a), where the \acs{ris} is used for target sensing in which the direct path may or may not be present (“\acs{ris}-aided”); setting (b), where the sensing is carried out solely using the path via the \acs{ris} (“\acs{ris} only”); and setting (c), where we do not have an \acs{ris} (“Without \acs{ris}”). For the latter two cases, the optimal solution can be computed in closed form. In  Figs.~\ref{fig:ris_aided_sensing2}(c) and ~\ref{fig:ris_aided_sensing2}(d), we illustrate the target illumination power and CRB for an \acs{ris} with $N=100$ elements. It is shown that, as the target is moving closer to the \acs{ris}, the strength of the path via the \acs{ris} increases, resulting in a larger target illumination power via the metasurface. For the points being close to the Tx (i.e., points A, B, C, and D), the strength of the direct path is significantly higher than that of the path via the \acs{ris}. Hence, the target illumination power of the “\acs{ris}-aided” system is comparable to that of the “Without \acs{ris}” system. Interestingly, for points closer to the \acs{ris} (i.e., E and F), the strength of the path via the \acs{ris} is also relatively high, resulting in that the “\acs{ris} aided” system exhibits improved target illumination power than that of the “Without \acs{ris}” one. The direct path is completely blocked for points G and beyond, making the settings “\acs{ris} only” and “\acs{ris}-aided” identical. When the direct path is blocked, the target illumination power of a system without a direct path is $0$ (or $-\infty$ in dB), which makes it impossible to carry out any kind of sensing. Similar conclusions can be drawn from the analysis of the CRB. As before, for points close to the Tx, the CRBs of the “\acs{ris}-aided” and “Without \acs{ris}” systems are comparable, whereas, for points with the direct path blocked, the CRB of “\acs{ris}-aided” system, is identical to that of the “Only \acs{ris}” one. Moreover, for a system without an \acs{ris}, which holds for the points G up to J, the target illumination power is $0$ resulting in the CRB to be infinity, indicating that no sensing is possible. 

In Figs.~\ref{fig:ris_aided_sensing2}(e) and ~\ref{fig:ris_aided_sensing2}(f), we consider a larger \acs{ris} with $N=400$ elements. As depicted, due to the increased number of meta-atoms, the array gain offered by the \acs{ris} becomes larger, thereby increasing the strength of the path via the surface. Interestingly, for point F, the strength of the path via the \acs{ris} becomes higher than the direct path, which is conveyed by the fact that the target localization power (CRB) of the “\acs{ris} only” system is higher (lower) than that of the “Without \acs{ris}” syste,. Finally, it is important to note that the performance of the “\acs{ris}-aided” system is always better than or equal to that of the  “Without \acs{ris}” system. Hence, it can be concluded that, by deploying a properly designed RIS, the achievable sensing performance of the system will never be deteriorated. An \acs{ris} allows us to carry out sensing in scenarios that are otherwise impossible, thereby indicating the potential of the \ac{ris} technology in sensing applications.

It can be concluded from this section's numerical example that, in sensing-only applications, \acp{ris} are mostly beneficial in overcoming harsh and \ac{nlos} signal propagation environments, enabling sensing capability in scenarios where it is otherwise impossible. Following the spirit of this investigation, we discuss in the following section how \acp{ris} can enhance ISAC performance. It will be interestingly shown that, when sensing is combined with communications via \ac{isac}, \acp{ris} provide a fundamental gain, which is not restrictive to \ac{nlos} conditions.

\begin{figure}
    \centering
    \includegraphics[width=0.6\columnwidth]{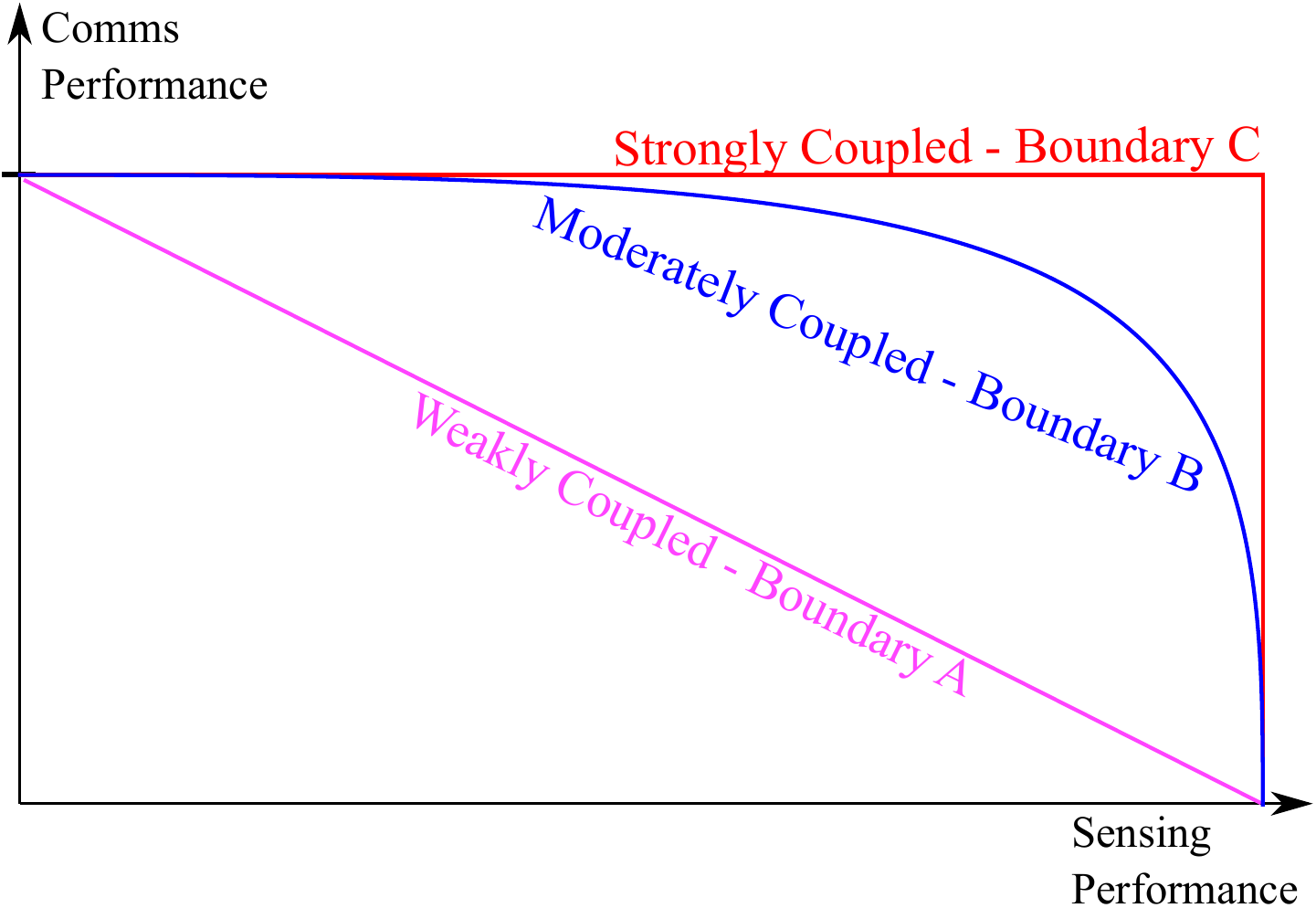}
    \caption{The Pareto boundaries for communications (comms) and sensing for three categories of the respective channels, examples of which are illustrated in Fig.~\ref{fig: SC channels correlation}.}
    \label{fig: ISAC_boundary}
\end{figure}

\section{\ac{ris}-Empowered \ac{isac}} \label{sec:ris-isac}
The potential of the RIS technology for either wireless communications \cite{CE_overview_2022} or sensing (primarily RF localization \cite{Keykhosravi2022infeasible}) is being lately widely investigated, indicating notable gains especially in scenarios where the direct Tx-Rx and Tx-target links are highly attenuated. Nevertheless, it still remains unclear if and how RISs can enhance \ac{isac} performance. To this end, a relevant fundamental question is: {\textit{What are the unique gains that RISs can provide to ISAC, in addition to those which have already been attained for their individual counterparts (i.e.,\ac{snc} functionalities)?}}

In this section, we attempt to answer the latter question from a high-level, yet practically relevant, viewpoint. 
In the sequel, we first showcase that the performance gain offer by ISAC systems over individual \ac{snc} ones originates from the coupling between the respective channels. We next explore the capability of RISs to manipulate this correlation, presenting design schemes to enhance the dual-functional performance of RIS-aided ISAC systems.

\subsection{Where Does the ISAC Gain Come From?}

\subsubsection*{\ac{snc} Trade-off}
A key challenge for the design of ISAC systems is the optimal design of the transmit signal waveform $\myVec{x}(t)$ in \eqref{eqn:Rx}, such that both the communications and sensing performances are improved. Since those individual performance metrics are rather different, there is, almost surely, an inherent performance trade-off between these functionalities. As a consequence, the ultimate goal for the $\myVec{x}(t)$ design is to reach, or approach, the Pareto frontier of some joint \ac{snc} metrics. In view of this perspective, a natural question arises: {{\textit{Do ISAC designs really provide any gains against individual communications or sensing systems?}}} The answer is intuitively affirmative since the wireless resources are shared between the two operations, however, a complete and accurate mathematical description is difficult to derive.

\begin{figure}
    \centering
    \includegraphics[width=0.5\columnwidth]{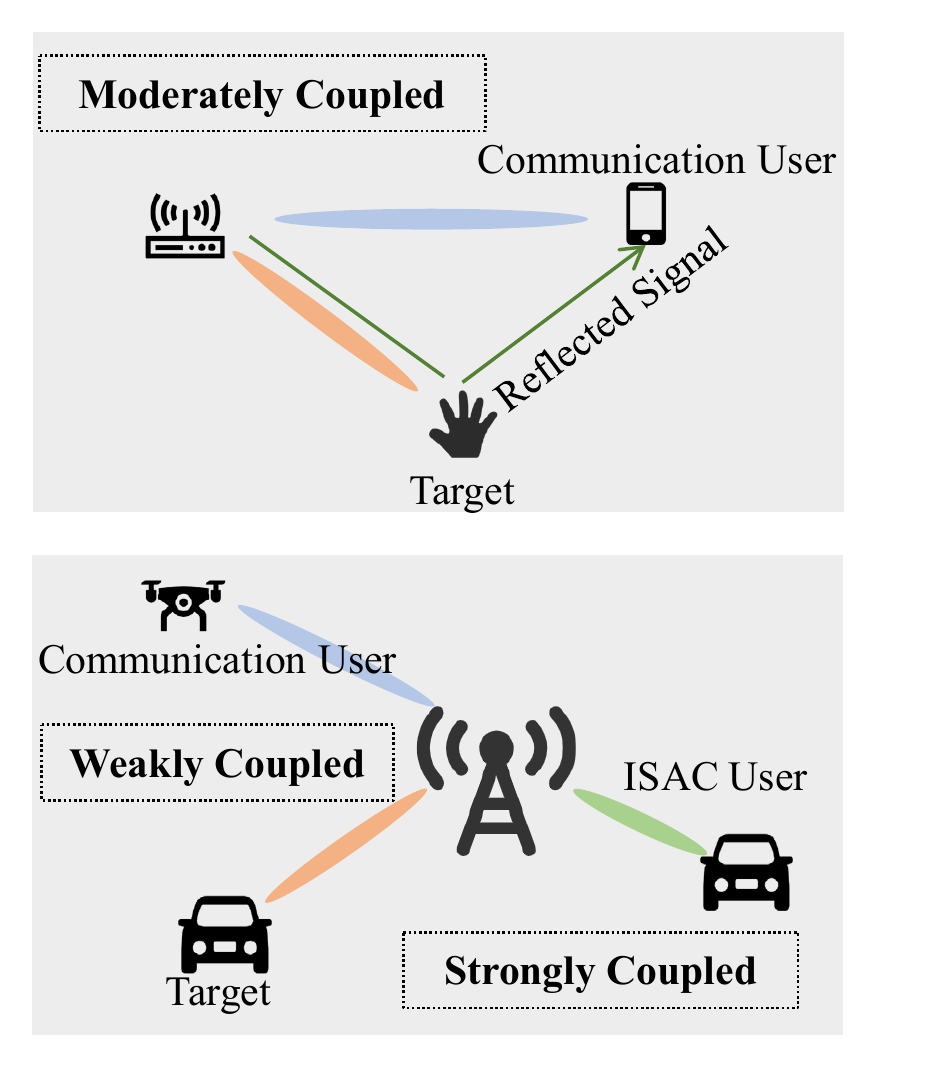}
    \caption{Three example categories of co-existing communications and sensing channels. }
    \label{fig: SC channels correlation}
\end{figure}

We illustrate a generic ISAC performance trade-off in Fig.~\ref{fig: ISAC_boundary}. Note that the figure is not constrained to specific performance metrics for communications or sensing; typical metrics are the channel capacity and detection probability, respectively. As shown, there exist in general three types of Pareto boundaries, namely, Boundaries A, B, and C. Intuitively, boundary C represents the best situation, i.e., both functionalities achieve always their optimal performance. On the other hand, Boundary A is the worst case, according to which the shared use of resources in ISAC provides negligible-to-zero gains compared to individual \ac{snc} systems. However, the most typical case is Boundary B, where the integration of \ac{snc} does provide performance gains over their individual consideration, but each functionality is unable to achieve its own optimum due to different design objectives.

\subsubsection*{\ac{snc} Coupling}
The boundaries A, B, and C in Fig.~\ref{fig: ISAC_boundary} correspond to three different ISAC scenarios, namely, the weakly-coupled, moderately-coupled, and strongly-coupled scenarios. Roughly speaking, a stronger coupling degree between the \ac{snc} channels, facilitates \acs{isac} systems with higher performance gains. We list the following examples to shed light onto this intuition, which are also illustrated in Fig.~\ref{fig: SC channels correlation}.
\begin{itemize}
    \item {\textbf{Weakly-Coupled Case:}} The communication channel is not correlated with the sensing one, e.g., the communication user is an unmanned aerial vehicle, while the target to sense is a ground vehicle.
    \item {\textbf{Moderately-Coupled Case:}} There exists partial correlation between the \ac{snc} channels, e.g., the sensing target is a hand gesture, which partially contributes to a wireless channel (e.g., in a WiFi channel).
    \item {\textbf{Strongly-Coupled Case:}} The communication channel is almost the same with the sensing channel, e.g., the sensing target is a vehicle, which is also as a communication user, thus requiring an ISAC system.
\end{itemize}

An illustrative example describing our high-level intuition for the importance of the \ac{snc} coupling on the \ac{isac} performance is provided in the box entitled {\it `ISAC Performance over \ac{snc} Channels of Varying Coupling Levels'} in the following page~\pageref{fig: SC channels correlation concept}. An analytical setup following this intuition is detailed in the sequel.
\begin{tcolorbox}[float*=t,
    width=\columnwidth,
	toprule = 0mm,
	bottomrule = 0mm,
	leftrule = 0mm,
	rightrule = 0mm,
	arc = 0mm,
	colframe = myblue,
	colback = mypurple,
	fonttitle = \sffamily\bfseries\large,
	title = ISAC Performance over \ac{snc} Channels of Varying Coupling Levels]
    \includegraphics[width=\columnwidth]{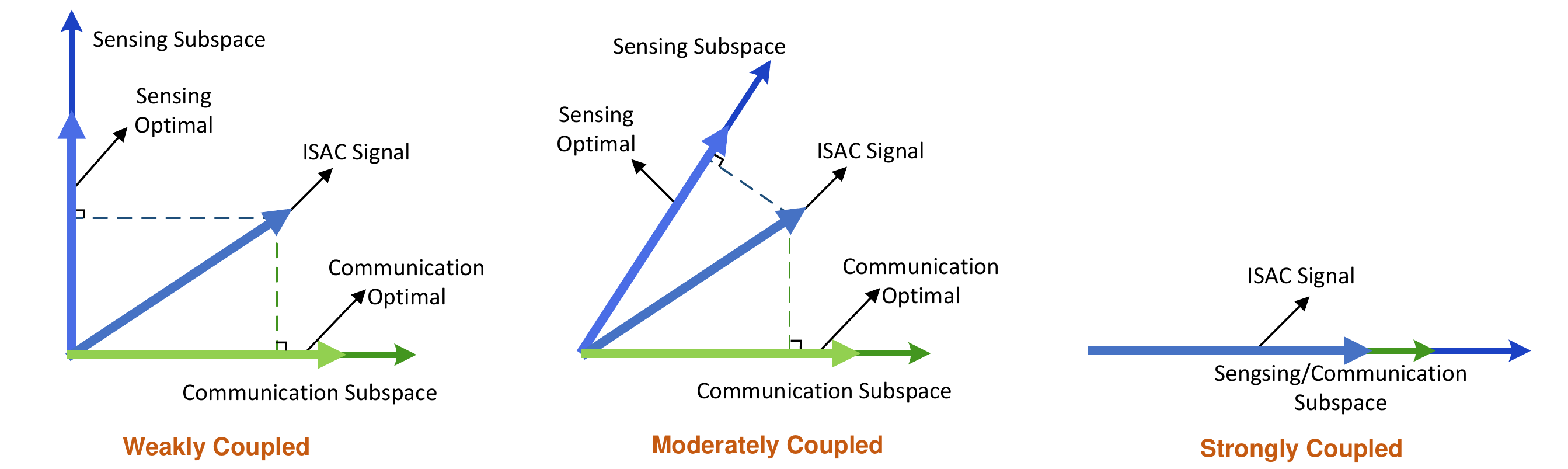}
    \label{fig: SC channels correlation concept}
    
This figures sketches the ISAC performance gains under different correlation levels for the co-existing \ac{snc} channels. In general, an ISAC signal should belong to the space spanned by the communications and sensing subspaces. The correlation between these two subspaces can be represented by their intersection angle. Accordingly, the \ac{snc} performance can be measured by projecting the ISAC signal onto each of the corresponding subspaces. In the weakly-coupled case, the two subspaces are nearly orthogonal to each other, which results close to the minimum projection. In the moderately-coupled case, the projection onto the sensing subspace becomes larger, given that the projection length on the communication subspace is the same. Finally, in the strongly-coupled case, the two subspaces are fully aligned to each other, and both the \ac{snc} performances achieve their optimum performances.
\end{tcolorbox}	

\subsubsection*{\ac{snc} Coupling Effect versus \ac{isac} Beamforming}
We consider a MIMO ISAC system to analytically show the importance of the \ac{snc} coupling and the ability of \ac{isac} systems in leveraging this property. An ISAC \ac{bs} equipped with a \ac{ula} is assumed to serve a single-antenna user, i.e., $\NRxCom=1$, while simultaneously being capable to detect a point-like target. In this case, the signals reflected back to the Rx by the target and received at the communications (comms) user are given respectively as follows:
\begin{equation} \label{motivating_model}
  \begin{gathered}
  \text{Sensing Model: }{\myVec{y}_s}\left( t \right) = \eta {\myVec{a}}_r\left( \theta  \right){{\myVec{a}}_t^T}\left( \theta  \right){\vw}s\left( t \right) + {{\myVec{z}}_s}\left( t \right), \hfill \\
  \text{Comms Model: }{y_c}\left( t \right) = {\myVec{h}}_c^T{\vw}s\left( t \right) + {z_c}\left( t \right),\hfill
  \end{gathered}
\end{equation}
where $\eta$ and $\theta$ are the amplitude and azimuth angle of the target, as well as $\myVec{a}_t\left(\theta\right) \in \mathbb{C}^{\NTx }$ and $\myVec{a}_r\left(\theta\right) \in \mathbb{C}^{\NRxRad }$ are the ULA transmit and receive steering vectors. In addition, $\myVec{h}_c \in \mathbb{C}^{\NTx }$ represents the communication channel between the BS and the single-antenna user. Moreover, ${\myVec{x}}\left( t \right) = {\vw}s\left( t \right)$ is an ISAC signal defined on the space-time domain, with ${\vw}\in \mathbb{C}^{\NTx }$ being the transmit beamformer and $s\left( t \right)$ denoting the communication data stream, which is assumed to be Gaussian distributed with zero mean and unit variance. Finally, $z_s(t)$ and $z_c(t)$ represent the white Gaussian noise terms of variance $\sigma_s^2$ and $\sigma_c^2$ at the communication and sensing receivers, respectively.

{\bf \ac{isac} Beamforming:}
Based on (\ref{motivating_model}), the communication performance can be measured through the achievable rate as
\begin{equation}\label{comms_rate}
    {R_c} = \log \left( {1 + {{{{\left| {{\myVec{h}}_c^T{\vw}} \right|}^2}}}/{{\sigma _c^2}}} \right).
\end{equation}
Without loss of generality, considering a conjugate symmetric receive array as in Section~\ref{sed:det}, the CRB for estimating the unknown target angle $\theta$ in \eqref{eq:crb} can be approximated (at large SNR values) as follows~\cite{van2004optimum}:

\begin{equation}\label{sensing_CRB}
\text{CRB}\left( \theta  \right) = \frac{{\sigma _s^2}\NRxRad}{{2T{\sigma_\eta^2}{{\left\| {{\mathbf{\dot a}}}_r \right\|^2_2}}{{\left| {{{\myVec{a}}_t^T}{\vw}} \right|}^2}}},
\end{equation}
where recall that $\sigma_\eta^2 = \mathbb{E}[{|\eta|^2}]$ and the term ${\left| {{{\myVec{a}}_t^T}{\vw}} \right|}^2$ represents the target illumination power. 

To serve the dual purpose of simultaneous \ac{snc}, one may formulate the following optimization problem for the design of the ISAC beamformer $\vw$:
\begin{equation}\label{opt_problem}
\begin{gathered}
  \underset{\vw}{\text{minimize}} \;\operatorname{CRB} \left( \theta  \right) \hfill \\
  \text{s.t.}\;\;\;{R_c} \ge {R_0},\;{\left\| {\vw} \right\|_2^2} \le 1, \hfill \\ 
\end{gathered}
\end{equation}
where $R_0$ denotes a communication rate performance threshold. We next minimize the CRB of the estimation for the target's angle, subject to the \ac{qos} constraint for downlink communications. By relying on (\ref{comms_rate}) and (\ref{sensing_CRB}), problem (\ref{opt_problem}) can be rewritten in compact form as follows:
\begin{equation}\label{simplified_opt_problem}
\begin{gathered}
  \underset{\vw}{\text{maximize}} \;{\left| {{{\myVec{a}}_t^T}{\vw}} \right|^2} \hfill \\
  \text{s.t.}\;\;\;{\left| {{\myVec{h}}_c^T{\vw}} \right|^2} \ge \sigma _c^2\left( {{2^{{R_0}}} - 1} \right),\;{\left\| {\vw} \right\|^2_2} \le 1. \hfill \\ 
\end{gathered}
\end{equation}
As a consequence, the ISAC beamforming design problem reduces to a simpler version, which necessitates the maximization of the radiation power towards the target's angle $\theta$, while guaranteeing the rate-achieving SNR theshold at the user's direction $\myVec{h}_c$. Intuitively, the optimal solution $\vw^\star$ for \eqref{simplified_opt_problem} trades off between the directions of the \ac{snc} channels. More precisely, it can be shown \cite{liu_CRB} that ${{\vw}^ \star } \in \operatorname{span} \left\{ {{\myVec{a}}_t,{{\myVec{h}}_c}} \right\}$:
\begin{equation}\label{closed_form}
  {{\vw}^\star} = \left\{ \begin{gathered}
   \frac{{\myVec{a}}_t}{{\left\| {\myVec{a}}_t \right\|_2}},\;\;{\text{if}}\;{\left| {{\myVec{h}}_c^T{\myVec{a}}}_t \right|^2} >N\left( {{2^{{R_0}}} - 1} \right)\sigma _c^2 \hfill \\
  {\lambda_1}{\tilde{\myVec{ h}}_c} + {\lambda_2}{\tilde{\myVec{ a}}},\;{\text{otherwise}} \hfill \\
\end{gathered}  \right.,
\end{equation}
where we have used the definitions:
\begin{equation}
\begin{gathered}
  {{\tilde{\myVec{ h}}}_c} = \frac{{{{\myVec{h}}_c}}}{{\left\| {{{\myVec{h}}_c}} \right\|_2}},\quad \tilde{\myVec{ a}} = \frac{{{\myVec{a}}_t - \left( {\tilde{\myVec{ h}}_c^T{\myVec{a}}_t^*} \right){{\tilde{\myVec{ h}}}_c}}}{{\left\| {{\myVec{a}}_t - \left( {\tilde{\myVec{ h}}_c^T{\myVec{a}}_t^*} \right){{\tilde{\myVec{ h}}}_c}} \right\|_2}},\hfill \\
  {\lambda_1} = \sqrt {\frac{{\left( {{2^{{R_0}}} - 1} \right)\sigma _c^2}}{{{{\left\| {{{\myVec{h}}_c}} \right\|_2^2}}}}} \frac{{\tilde{\myVec{ h}}_c^T{\myVec{a}}_t^*}}{{\left| {\tilde{\myVec{ h}}_c^T{\myVec{a}}_t^*} \right|}},\hfill \\ {\lambda_2} = \sqrt {{P_T} - \frac{{\left( {{2^{{R_0}}} - 1} \right)\sigma _C^2}}{{{{\left\| {{{\myVec{h}}_c}} \right\|_2^2}}}}} \frac{{{{\tilde{\myVec{ a}}}^T}{\myVec{a}}_t^*}}{{\left| {{{\tilde{\myVec{ a}}}^T}{\myVec{a}}_t^*} \right|}}. \hfill \\ 
\end{gathered} 
\end{equation}

{\bf Channel Coupling Effect:}
By inspecting (\ref{closed_form}), we observe that the performance trade-off between the \ac{snc} operations can be essentially described by the process that moves the ISAC signal from the sensing subspace $\operatorname{span} \left\{ {{\myVec{a}}_t} \right\}$ to the communication subspace $\operatorname{span} \left\{ {{{\myVec{h}}_c}} \right\}$. Apparently, the intersection angle between these two subspaces determines the coupling/correlation degree of the two channels, and, accordingly, regulates the ISAC performance gain over the individual \ac{snc} functionalities. We examine the following two special cases:
\begin{itemize}
    \item \textbf{Fully-Aligned Subspaces:} In this case, the \ac{snc} channels are strongly correlated, and the optimal beamformer is $\vw^\star =  \frac{{\myVec{a}}_t}{{\left\| {\myVec{a}}_t \right\|_2}} = \frac{{\myVec{h}}_c}{{\left\| {\myVec{h}}_c \right\|_2}}$, which generates the following CRB and achievable rate:
    \begin{equation}\label{strong_correlated_CRB}
      {\text{CRB}}\left( \theta  \right) = \frac{{\sigma _s^2}\NRxRad}{{2{{\sigma_\eta}^2}T{{\left\| {{\mathbf{\dot a}}}_r \right\|_2^2}}{{\left\| {\myVec{a}}_t \right\|_2^2}}}},
    \end{equation}
    \begin{equation}\label{strong_correlated_rate}
        {R_c} = \log \left( {1 + {{{{\left\| {{{\myVec{h}}_c}} \right\|_2^2}}}}/{{\sigma _c^2}}} \right). 
    \end{equation}
    This implies that, for any feasible rate $R_0$ being smaller or equal to $R_c$ in (\ref{strong_correlated_rate}), both the \ac{snc} functionalities achieve their best performance without any trade-off. In other words, the system's resources can be fully reused between \ac{snc} and the ISAC gain is maximized.
    \item \textbf{Orthogonal Subspaces:} In this case, the two channels are decoupled, i.e., ${\myVec{h}}_c^H{\myVec{a}}_t = 0$, and $R_c = R_0$. The CRB can be thus expressed as a function of $R_0$ (or $R_c$) as:
    \begin{equation}\label{zero_correlation_CRB_rate}
        {\text{CRB}}\left( \theta  \right) = \frac{{\sigma _s^2}\NRxRad}{{2{{\sigma_\eta}^2}T\left( {1 - \frac{{\left( {{2^{{R_0}}} - 1} \right)\sigma _c^2}}{{{{\left\| {{{\myVec{h}}_c}} \right\|}^2}}}} \right){{\left\| {{\mathbf{\dot a}}}_r \right\|}^2_2}{{\left\| {\myVec{a}}_t \right\|}^2_2}}}.
    \end{equation}
    By comparing (\ref{zero_correlation_CRB_rate}) with (\ref{strong_correlated_CRB}), we observe that the \ac{crb} performance is increased due to the \ac{qos} requirement of the communication user. In this case, the ISAC signal has to be decomposed into two orthogonal directions, with no resources being reused. Therefore, there is no ISAC gain for orthogonal \ac{snc} subaspaces, since the resource efficiency is the same as that of the individual \ac{snc} functionalities.
\end{itemize}

In the general cases where the two \ac{snc} subspaces are neither aligned nor orthogonal, part of the signal power can be shared between the two operations, and the remaining part has to be decomposed into two orthogonal directions. The ISAC gain is thus determined by the portion of the reused signal power, which depends on the intersection angle between the \ac{snc} subspaces.
\begin{figure}
    \centering
    \includegraphics[width=0.6\columnwidth]{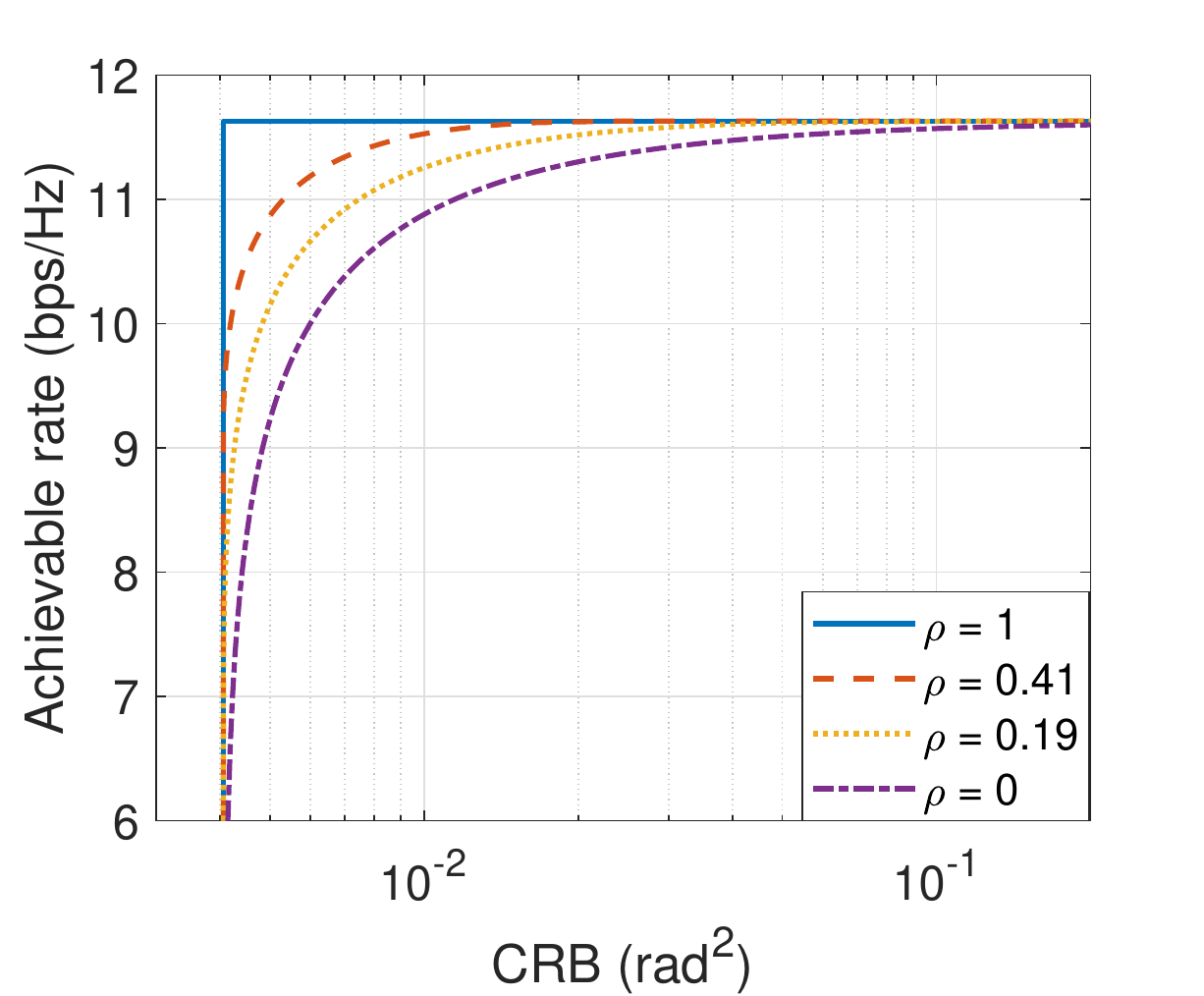}
    \caption{Sensing and communication performance trade-off under different values for the correlation coefficient $\rho$.}
    \label{fig: SC channels correlation2}
\end{figure}
To showcase this behaviour, we present a numerical example for the dual-operation performance in Fig.~\ref{fig: SC channels correlation2}, considering the setting of parameters included in Table~\ref{tab:my_label}. In particular, the \ac{snc} performance trade-off is illustrated under different correlation coefficient between the ac{snc} subspaces, which is defined as $\rho  = \frac{{\left| {{\myVec{h}}_c^H{\myVec{a}}_t} \right|}}{{\left\| {{{\myVec{h}}_c}} \right\|_2\left\| {\myVec{a}}_t \right\|_2}}$. It can be observed that the increase of the correlation between the \ac{snc} channels yields significant ISAC performance gains.
\begin{table}[]
\centering
\caption{Parameters for the Results in Fig.~\ref{fig: SC channels correlation2}.}
\label{tab:my_label}
{
\begin{tabular}{c c}
\hline 
\textbf{Parameter} & \textbf{Value}\\
\hline
Transmit Power  & 1W \\
$\sigma_s^2$ & -60 dBm \\
$\sigma_c^2$ & -60 dBm \\
Center Frequency & 3 GHz \\
$\NTx$ & 15  \\
$\NRxRad$ & 15  \\
Location of the BS & [0,0] m \\
Location of the Target & [40,0] m \\
\hline
\end{tabular}
}
\end{table}
 
%
\subsection{How Do RISs Improve ISAC Performance?}
\label{ssec:ISAC2}
\begin{tcolorbox}[float*=t,
    width=\columnwidth,
	toprule = 0mm,
	bottomrule = 0mm,
	leftrule = 0mm,
	rightrule = 0mm,
	arc = 0mm,
	colframe = myblue,
	colback = mypurple,
	fonttitle = \sffamily\bfseries\large,
	title = ISAC Performance Improvement via RISs: Subspace Expansion and Rotation]
	\label{box:ISAC_w_RIS}
    \begin{center}
    \includegraphics[width=0.5\columnwidth]{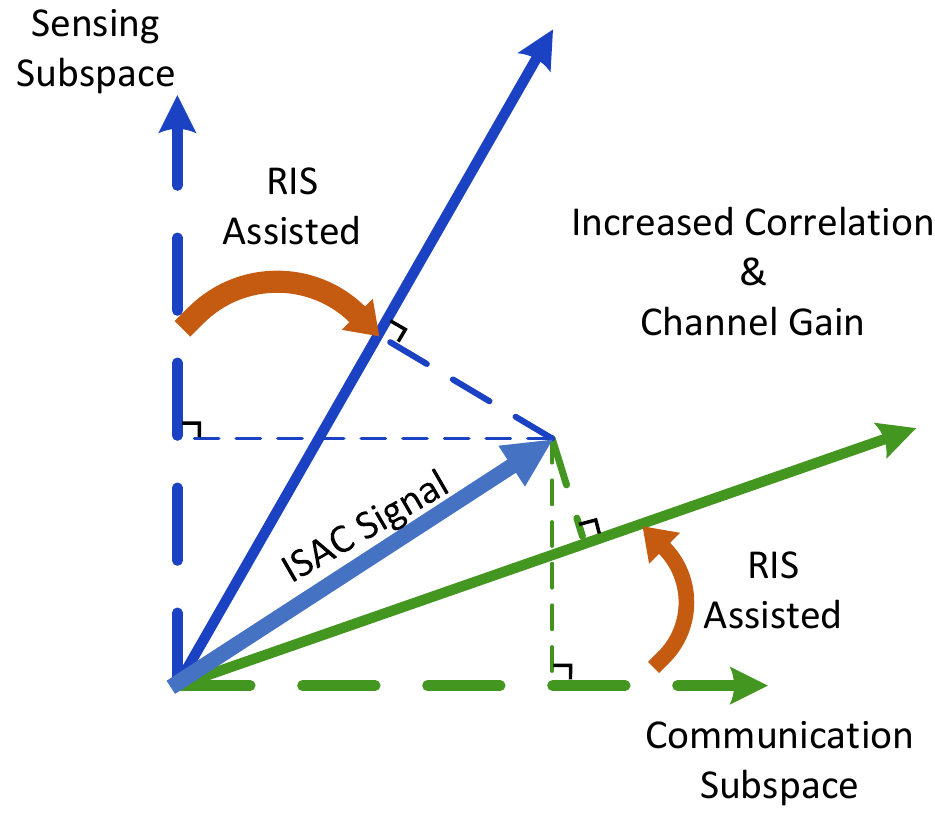}
    \end{center}
    
The ISAC performance gain provided by an RIS is schematically demonstrated in the figure on the left for weakly-coupled \ac{snc} channels \cite{Meng_arxiv}. In general, RISs are capable of offering two positive effects for ISAC operation. First, an RIS provides additional channel paths, and consequently channel gains, for both \ac{snc}, which is equivalent to expanding the respective subspaces. Second, by optimizing the phases of the RIS, the previously nearly orthogonal \ac{snc} subspaces can be rotated to become coupled/correlated between each other. As a consequence, RISs can significantly improve the ISAC trade-off performance.
\end{tcolorbox}	

In the previous subsections, we presented the relationship between the \ac{snc} channel coupling and the gains from the dual-functional design via \ac{isac}.
We now detail the gains that \acp{ris} can provide to ISAC systems, in light of the above insight. The high-level rationale of this subsection is included in the box entitled {\it `ISAC Performance Improvement via RISs: Subspace Expansion and Rotation`} in the next page~\pageref{box:ISAC_w_RIS}. In addition, we provide an analytical example for the provided gains. To keep our analysis simple and tractable, we focus on the conventional model of \ac{ris}-aided channels according to which, the \ac{ris} operation is represented by controllable phase profiles and the overall channel obeys the cascaded channel model, e.g., as presented in \eqref{eqn:Cascaded}. In particular, we show that, by manipulating the RIS phase configuration, the correlation between the \ac{snc} channels can be improved, which can yield additional ISAC gains. 

{\bf \ac{isac} Beamforming with RISs:}
Similar to before, we focus on the \ac{isac} beamforming design problem, considering now the integration of an RIS with $\NRis$ meta-atoms, which is intended to improve the joint \ac{snc} performance. To this end, the \ac{snc} signal models of \eqref{motivating_model} can now be re-expressed as:
\begin{equation} \label{motivating_model_RIS}
\begin{gathered}
  {\text{Sensing Model: }}{\myVec{y}_s}\left( t \right) = {{\myVec{h}}_r}{\myVec{h}}_t^T{\vw}s\left( t \right) + {{\myVec{z}}_s}\left( t \right) \hfill \\
  {\text{Comms Model: }}{y_c}\left( t \right) = {\myVec{h}}_c^T{\vw}s\left( t \right) + {z_c}\left( t \right) \hfill \\ 
\end{gathered}
\end{equation}
where the channel gains ${{\myVec{h}}_r}$ and ${{\myVec{h}}_t}$ are defined in (\ref{eq:RIS_channels}) and
\begin{equation}\label{RIS_comms_channels}
\begin{gathered}
  {{\myVec{h}}_c} = {{\myVec{h}}_{\rm BU}} + {{\mathbf{G}}_t}{\mathbf{\Phi }}{{\myVec{h}}_{\rm RU}}. \hfill \\ 
\end{gathered}
\end{equation}
In the latter expression, $\myVec{h}_{\rm BU} \in \mathbb{C}^{\NTx }$ and $\myVec{h}_{\rm RU} \in \mathbb{C}^{\NRis}$ represent the direct channel and the channel between the RIS and the communication Rx, respectively, with the latter being henceforth referred to as the \ac{ue}. In this model, a dual-functional signal ${\vx}\left( t \right)={\vw}s\left( t \right)$ is transmitted through the ISAC BS, and is reflected by the RIS. The signals received at both the \ac{ue} and the target are from two propagation paths, i.e., the direct path from the BS, and the reflective path from the RIS. Again, we denote by ${y_c}\left( t \right)$ the signal received at the \ac{ue}, and the target is assumed to reflect back echo signals to the ISAC BS through both paths. The resulting sensing signal model ${\myVec{y}_s}\left( t \right)$ is provided in (\ref{motivating_model_RIS}).

{\bf RIS-Aided Channel Coupling Effect:}
By inspecting (\ref{motivating_model_RIS}), we can obtain the following observations:
\begin{enumerate}[label=({\it O\arabic*})]
    \item \label{itm:Obs1} For both \ac{snc} functionalities, the presence of the RIS provides extra signal propagation paths. This property enhances the channel gain for both target sensing and downlink data communications.
    \item \label{itm:Obs4} The adjustable phase profiles of the RIS offer another unique and promising way to improve the ISAC performance. This is capability to artificially increase the correlation between the \ac{snc} channels, such that more signal power can be reused by the dual functionalities to boost the ISAC efficiency. 
\end{enumerate}

\begin{figure}
    \centering
    \includegraphics[width=0.6\columnwidth]{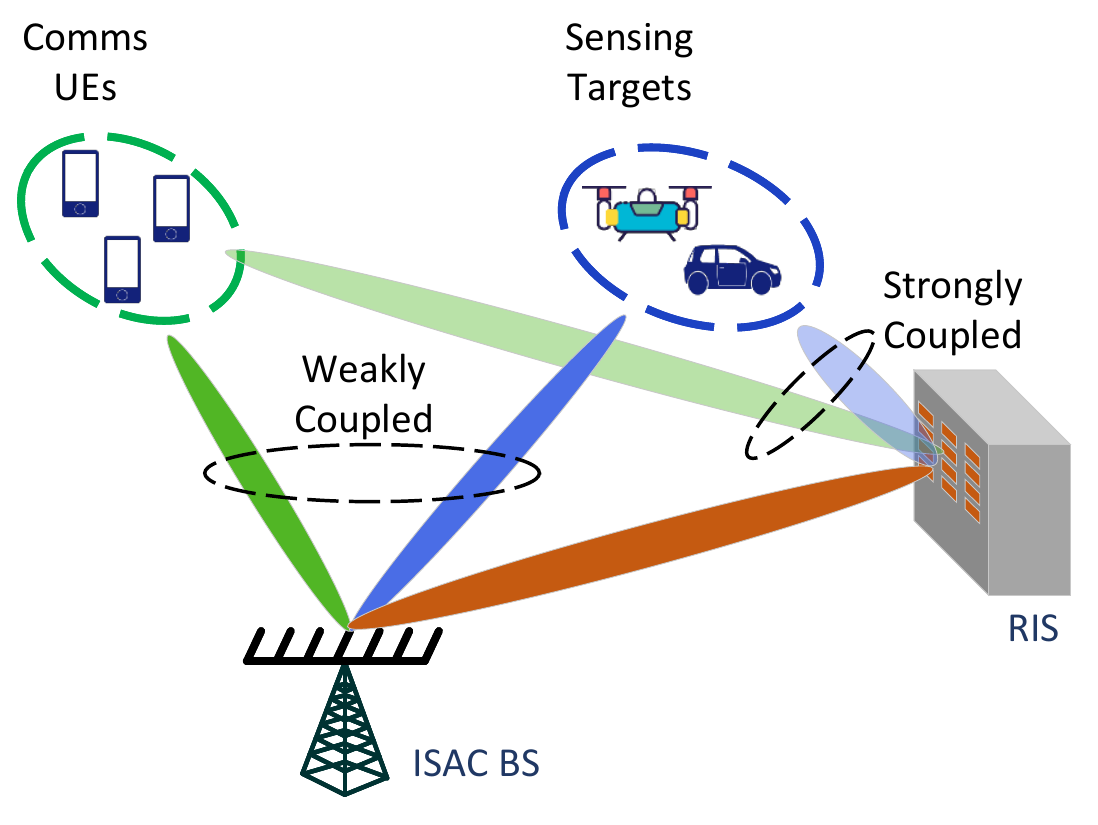}
    \caption{An RIS-aided ISAC system illustrating the observation~\ref{itm:Obs4}; the figure has been adopted from \cite{Meng_arxiv}.}
    \label{fig: SC channels correlation RIS}
\end{figure}

We illustrate the latter observation \ref{itm:Obs4} in Fig.~\ref{fig: SC channels correlation RIS}, where two groups of communications \acp{ue} and sensing targets are well separated, resulting in weakly coupled direct channels. By placing an \acs{ris} at a proper position and tuning its reflection pattern accordingly, strongly coupled beams can be formulated by the \acs{ris} towards the \acp{ue}' and the targets' clusters. As a consequence, the resulting \ac{snc} channels become moderately coupled, enhancing the gains of the \ac{isac} operation.

%
\subsection{RIS-Aided ISAC Beamforming based on Subspace Expansion and Rotation}
The observations \ref{itm:Obs1} and \ref{itm:Obs4} were formulated for a specific setting of the RIS-aided \ac{isac} beamformer, considering a single-antenna \ac{ue}. Nonetheless, they reveal the gain offered by \ac{ris}-aided \ac{isac} systems in a holistic fashion, which stems from the capability of \acp{ris} to enhance the level of coupling between the \ac{snc} channels, and thus improve the underlying trade-off in the \ac{isac} operation. This capability relies on dedicated signal processing techniques for tuning the RIS phase profiles along with the dual-function transmission schemes. 
We next present a two-step beamforming design aiming to improve the simultaneous \ac{snc} performance. The basic idea is to first maximize the correlation between the \ac{snc} channels' correlation and gains, by manipulating the RIS phase profile to expand and rotate the respective subspaces~\cite{Meng_arxiv}. Then, with the favorable \ac{snc} channel characteristics, the transmit beamformer at the BS can be further optimized with the goal to minimize the estimation CRB, while ensuring a desirable communication rate. 

We first rewrite the sensing signal model in (\ref{motivating_model_RIS}) as follows:
\begin{equation}\label{RIS_sensing_rewrite}
    {\myVec{y}_s}\left( t \right) = \beta {\myMat{H}}\left( {\bm{\theta }} \right){\vw}s\left( t \right) + {{\myVec{z}}_s}\left( t \right),
\end{equation}
where $\beta = \alpha_r\alpha_t$, ${\myMat{H}}\left( {\bm{\theta }} \right) = {{{{\myVec{h}}_r}{{\myVec{h}}_t}} \mathord{\left/
 {\vphantom {{{{\myVec{h}}_r}{{\myVec{h}}_t}} \beta }} \right.
 \kern-\nulldelimiterspace} \beta }$, and $\bm \theta = \left[\theta_1,\theta_2\right]^T$. To ease the derivation, we assume that the parameters to be estimated are $\beta$ and $\bm \theta$, while all other parameters are assumed to be fixed. Then, we formulate the following subspace expansion and rotation problem:
 \begin{equation}
 \label{correlation_problem}
\begin{gathered}
  \underset{{\mathbf{\Phi }}}{\text{maximize}} \;{\left\| {{{\myMat{H}}^H}{{\myVec{h}}_c}} \right\|_2^2} \hfill \\
  \text{s.t.}\;\;\;\left| {{\phi _i}} \right| = 1, i = 1,, \ldots ,\NRis,
\end{gathered}
 \end{equation}
As can be seen, the objective function is the inner product of the \ac{snc} channels, which represents both the channel gains and the \ac{snc} correlation. By defining ${\bm{\phi}} = {\left[ {{\phi _1},{\phi _2}, \ldots ,{\phi _\NRis}} \right]^T}$, the latter optimization problem can be recast as:
\begin{equation}\label{recast_correlation_problem}
\begin{gathered}
  \underset{\bm\phi}{\text{minimize}}  \;\;f\left(\bm \phi\right) \hfill \\
  \text{s.t.}\;\;\;\left| {{\phi _i}} \right| = 1, i = 1,, \ldots ,\NRis,
\end{gathered}  
\end{equation}
where we have used the function definition:
\[f\left(\bm \phi\right) = -{\left\| {{{\myVec{a}}_t} + {{\mathbf{F}}_t}{\bm\phi} } \right\|_2^2}{\left| {{{\left( {{{\myVec{a}}_r} + {{\mathbf{F}}_r}{\bm\phi} } \right)}^H}\left( {{{\myVec{h}}_{BU}} + {{\mathbf{F}}_c}{\bm\phi} } \right)} \right|^2}.\] with ${{\mathbf{F}}_t} = {{\mathbf{G}}_t}\operatorname{diag} \left\{ {\myVec{b}} \right\}$, ${{\mathbf{F}}_r} = {{\mathbf{G}}_r}\operatorname{diag} \left\{ {\myVec{b}} \right\}$, and ${{\mathbf{F}}_c} = {{\mathbf{G}}_t}\operatorname{diag} \left\{ {{{\myVec{h}}_{\rm RU}}} \right\}$. It can be easily concluded that the problem (\ref{recast_correlation_problem}) is non-convex due to both the non-convex objective function and the unit-modulus constraint on $\bm \phi$. However, one may readily employ a gradient descent method to seek for a local optimum, followed by projection, which is simply to normalize each element to have unit modulus. 


\begin{figure}
    \centering
    \includegraphics[width=0.6\columnwidth]{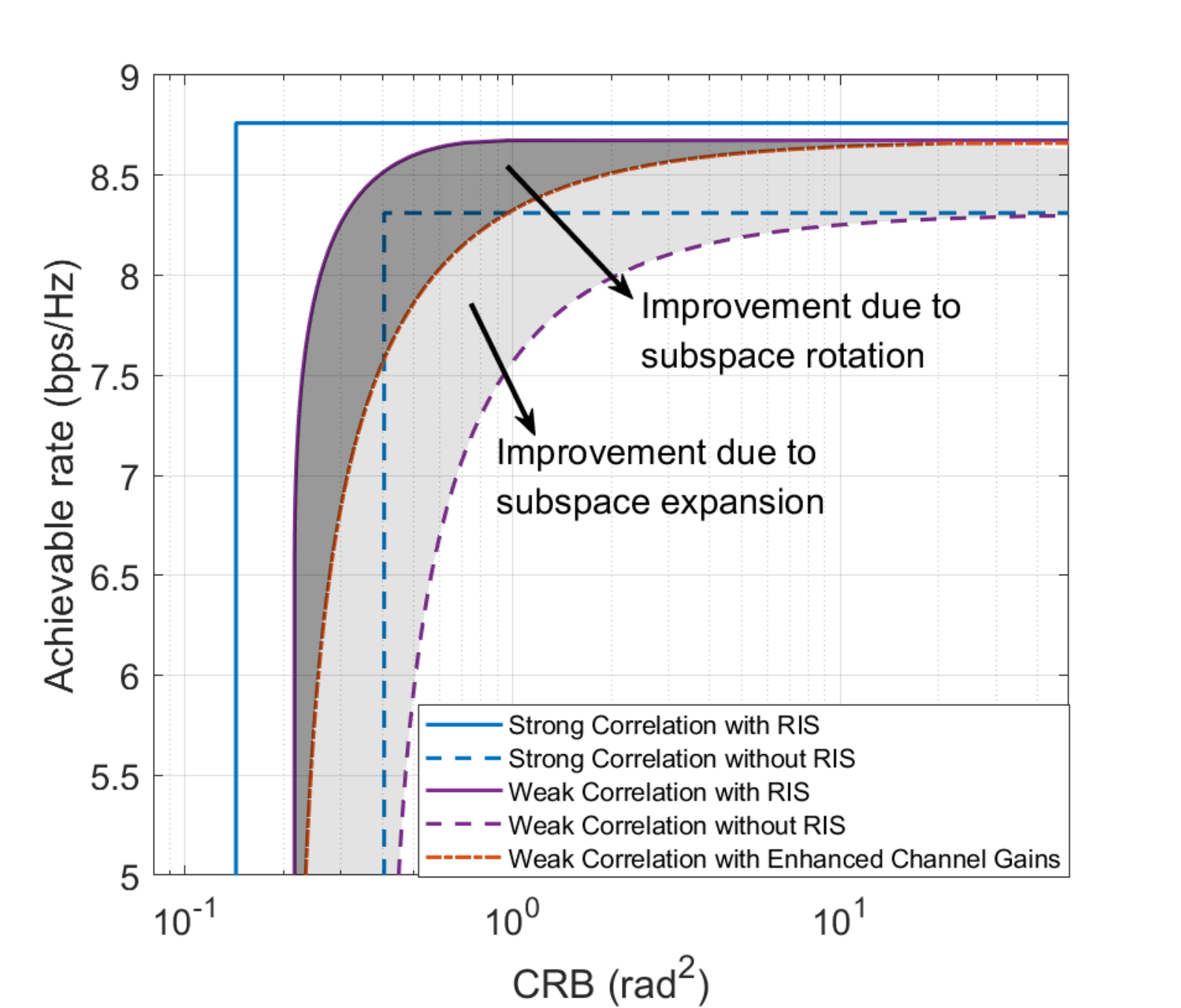}
    \caption{ISAC performance with and without an RIS for both the strongly- and weakly-coupled cases; the figure has been adopted from \cite{Meng_arxiv}.}
    \label{fig: SC channels correlation RIS2}
\end{figure}

The next step is to design a dual-functional beamformer $\vw$ by solving the following optimization problem:
\begin{equation}\label{RIS_ISAC_BF_design}
\begin{gathered}
  \underset{\vw}{\text{minimize}} \;\operatorname{CRB} \left( {\theta_1}  \right) \hfill \\
  \text{s.t.}\;\;\;\log \left( {1 + {{{{\left| {{\myVec{h}}_c^T{\vw}} \right|}^2}}}/{{\sigma _c^2}}} \right) \ge {R_0}, 
  \quad{\left\| {\vw} \right\|^2} \le {1}. \hfill \\ 
\end{gathered}
\end{equation}
Similar to the non-RIS case, we here minimize the CRB of the estimation of the target's \ac{aoa} relative to the BS's antenna array, subject to a communication rate constraint. It is noted that the CRB of $\theta_1$ can be obtained from the computation of the inverse of the Fisher Information Matrix (FIM), followed by the selection of its corresponding diagonal entries. This Tx beamforming design problem is again a non-convex optimization problem, which can be solved using semi-definite relaxation (SDR) methods.
\begin{table}[]
\centering
\caption{Parameters for the results in Fig.~\ref{fig: SC channels correlation RIS2}}
\label{tab:my_label2}
{
\begin{tabular}{c c}
\hline 
\textbf{Parameter} & \textbf{Value}\\
\hline
Transmit Power  & 3W \\
$\sigma_s^2$ & -50 dBm \\
$\sigma_c^2$ & -50 dBm \\
Center Frequency & 3 GHz \\
$\NTx$ & 15  \\
$\NRxRad$ & 15  \\
$\NRis $ & 64  \\
Location of the BS & [0,0] m \\
Location of the Target & [40,0] m \\
Location of the RIS & [30,30] m \\
\hline
\end{tabular}
}
\end{table}

In Fig.~\ref{fig: SC channels correlation RIS2}, we present a numerical example for the above design methodology, considering both weakly- and strongly-coupled \ac{snc} channels. The setting of the simulation parameters for this figure are summarized in Table~\ref{tab:my_label2}. 
As depicted, in the strongly-coupled case, where the correlation coefficient of the \ac{snc} channels was set to $1$, a rectangular trade-off curve is observed irrespective of the RIS deployment. This actually suggests that this deployment improves the ISAC performance by simply enhancing the channel gains, i.e., by expanding the \ac{snc} subspaces. However, in the weakly-coupled case, the RIS benefit is more pronounced. To distinguish between the gains of subspace expansion and rotation, we have also drawn a reference curve with zero \ac{snc} channel correlation, but enhanced channel gains, such that the maximum achievable rate and the minimum CRB are made the same with those of the RIS-aided case. The origins of the improved trade-offs are highlighted in Fig.~\ref{fig: SC channels correlation RIS2}, where the light grey area depicts the performance gain provided by the enhanced channel gains, i.e., by subspace expansion, while the dark grey area characterizes the performance gain provided by the improved \ac{snc} channel correlation, i.e., by subspace rotation. This result demonstrates the effectiveness of the proposed RIS-aided ISAC beamforming approach for the considered scenarios. 

The above numerical findings indicate that, on top of the performance gains provided to the individual \ac{snc} systems, namely, NLoS localization and improved wireless throughput, RISs provide another unique opportunity for boosting the {\it {joint \ac{snc}}} in ISAC systems. This opportunity is shaped around their capability to control the coupling degree between the sensing and communication channels. It is noted that, although our derivation focuses only on a simple and analytically tractable system and channel model, it already offers new insights into the fundamental gains of RIS-empowered ISAC systems.

\subsection{RIS-Aided ISAC Beamforming based on Beampattern Errors}\label{sec:err}
\begin{figure*}
    \centering
    \includegraphics[width=\columnwidth]{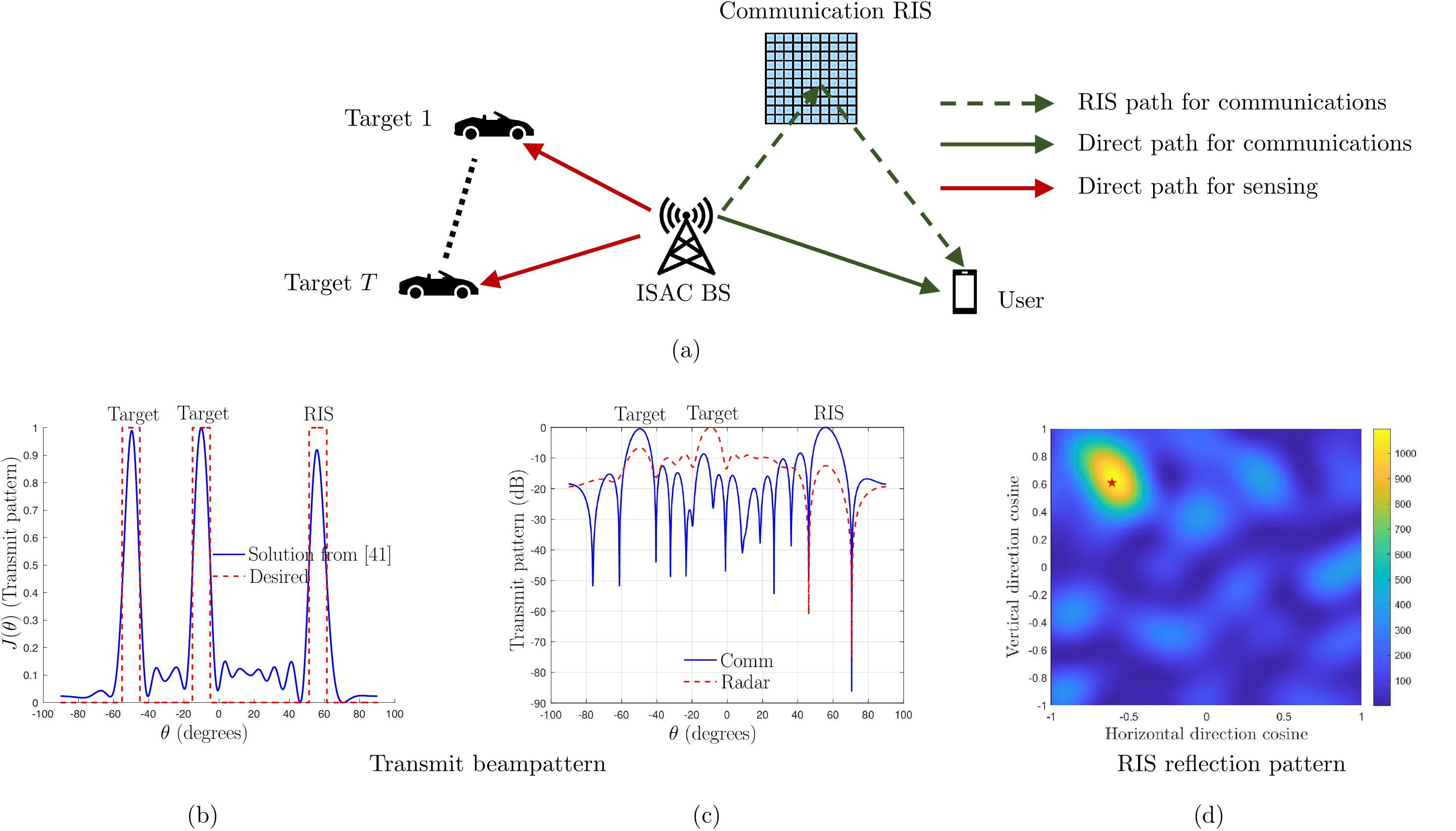}
       \caption{Transmit/reflection beamforming patterns, as have been adopted from \cite{sankar2022beamforming}, for the considered RIS-aided sensing system in Section~\ref{sec:err}. (a) The ISAC BS sends the superposition of precoded communication symbols and a sensing signal. (b) Transmit beampatterns towards the desired directions. (c) Individual communication and sensing beampatterns. (d) RIS reflection pattern. The symbol {\color{red}$\star$} indicates the user location with respect to the RIS.}
    \label{fig:dfbs_beampattern}
\end{figure*}

A sensing system usually requires waveforms with good cross-correlation properties, while being capable of matching a desired beampattern to focus energy towards directions of interest. In general, it is difficult to ensure such good properties for signals radiating from \acp{ris}, since they are expected to be highly correlated. To this end, one may consider to simultaneously transmit precoded communications and sensing waveforms through an ISAC BS, where the \ac{ris} can be deployed to only improve the communication performance (e.g., to improve the effective rank of the communication channel \cite{WavePropTCCN}) without affecting the correlation property of the sensing waveforms. This is illustrated in the ISAC system setup in Fig.~\ref{fig:dfbs_beampattern}(a), where the \ac{ris} is not used for sensing, but only for assisting communications to the user.

{
Consider the setting in Fig.~\ref{fig:dfbs_beampattern}(a) with $T$ targets and a single communication user, where the transmit signal is a superposition of a sensing waveform with precoded communication symbols, and is expressed as follows:
\[
\myVec{x}\left( t \right) = \vc d\left( t \right) + {\mW}\vs\left( t \right),
\]
where  $\vc\in\mathbb{C}^{L_{\rm T}}$ is the communication precoder, $d(t)$ is the communication symbol transmitted to the user at time $t$, and $\mW\in\mathbb{C}^{L_{\rm T}\times T}$ is the sensing beamformer that precodes the $T$ sensing signals in vector $\vs(t)$ at time $t$. Assuming that the sensing signals and communication symbols are uncorrelated with each other both having unit power, i.e., $\mathbb{E}[|d(t)|^2] = 1$, $\mathbb{E}[\vs(t)\vs^H(t)] = {\bf I}$, and $\mathbb{E}[\vs(t)\bar{d}(t)] = {\bf 0}$, we can express the transmit covariance matrix as:
\[
\mR = \mathbb{E}[\myVec{x}(t)\myVec{x}^H(t)] = \vc\vc^H + \mW\mW^H.
\]
The signal received at the communication user in \eqref{motivating_model_RIS} now becomes as follows:
\[
{y_c}\left( t \right) = {\myVec{h}}_c^T[\vc d\left( t \right) + {\mW}\vs\left( t \right)] + {z_c}\left( t \right). 
\]
Recall that the channel ${\myVec{h}}_c$ defined in \eqref{RIS_comms_channels} depends on the \acs{ris} phase profile ${\boldsymbol \Phi}$. Then, the SINR (where the interference is due to the sensing waveform) at the user is expressed as:
	\begin{align}
		\gamma({\boldsymbol \Phi},\vc,\mR) 
		=\frac{ {\myVec{h}}_c^H  \vc \vc^H {\myVec{h}}_c }{{\myVec{h}}_c^H \left( \mR - \vc\vc^H \right){\myVec{h}}_c + \sigma_c^2}.
		 \label{sinr_k}
	\end{align}
	
The desired beampattern, denoted by $m(\theta)$, is a beampattern that ensures that sufficient power reaches the directions of the targets of interest and the communication \acs{ris} (see Fig.~\ref{fig:dfbs_beampattern}(a)). It could be a superposition of multiple rectangular beams, as shown in Fig.~\ref{fig:dfbs_beampattern}(b). The power radiated from the ISAC BS towards a direction $\theta$ is $J(\theta) = \mathbb{E}[|\myVec{a}_t^H(\theta)\myVec{x}(t)|^2] = \myVec{a}_t^H(\theta) \mR \myVec{a}_t(\theta).$ Given a desired beampattern $m(\theta)$, we may measure the sensing performance by the weighted combination of the beampattern mismatch error and the average cross-correlation of the signals reflected back by $T$ targets, i.e.:
\begin{align*}
		L(\mR,\tau) &= \alpha_1\frac{1}{D}\sum_{\ell=1}^{D} \vert J(\tilde{\theta}_{\ell}) - \tau m(\tilde{\theta}_{\ell})  \vert^2 \nonumber\\ &\quad+ \alpha_2 \frac{2}{T^2 - T}\sum_{i=1}^{T-1} \sum_{j=i+1}^{T} \vert 	\va_t^H(\tilde{\theta}_i)\mR \va_t^H(\tilde{\theta}_j) \vert^2,
\end{align*}
where $\tau$ is an unknown autoscale parameter, whereas $\alpha_1$ and $\alpha_2$ are known weights. This error is evaluated over a discrete grid of $D$ angles  $\{\tilde{\theta}_i\}_{i=1}^{D}$ and the second term in the above expression is the average squared cross-correlation. To design waveforms with good cross-correlation and desired beampatterns with good communication SINR performance, the following optimization problem was considered in~\cite{sankar2022beamforming}:
			\begin{align} 
				\quad \underset{\vc,\mW,\tau,\boldsymbol{\Phi}}{\text{minimize}} & \quad L(\mR,\tau)  \nonumber \\
				\text{s.t.} & \quad \mR = \vc\vc^H + \mW\mW^H \succeq {\bf 0}  \nonumber\\
				& \quad [\mR]_{i,i} = 1, \,\, i=1,\ldots,\NTx   \nonumber \\
				& \quad \gamma(\boldsymbol{\Phi},\vc,\mR) \geq \Gamma,  \nonumber\\
  & \quad\left| {{\phi _m}} \right| = 1,\quad  m = 1, \ldots ,\NRis, \hfill  
\end{align}
where we have normalized the transmit power and $\Gamma$ denotes the desired SINR. Clearly, this problem is non-convex in the optimization variables and suboptimal solvers based on alternating minimization were presented in~\cite{sankar2022beamforming}, together with an extended design formulation for a multi-user setting with different SINR thresholds for each user. 

We investigate the above joint design methodology via a numerical example, whose performance evaluation is illustrated in Figs.~\ref{fig:dfbs_beampattern}(b)-(d). In the former figure, the realized beampattern towards two target directions and the \acs{ris} are included. The individual sensing and communication beampatterns are shown in Fig.~\ref{fig:dfbs_beampattern}(b), where we can see that the communication beamformer has a stronger peak towards the \acs{ris} and the sensing beampattern implements a dip towards the \acs{ris}. This happens because if not, sensing waveforms would be transmitted towards the \acs{ris}, resulting in increased interference at the user side. On the other hand, transmitting communication symbols to targets improves the targets' illumination power. In Fig.~\ref{fig:dfbs_beampattern}(c), the phase profile of the \acs{ris} is depicted, where it is shown that the peak occurs at the user direction.

	
}


\section{Future Research Directions}
\label{sec:future_research}
Both \acp{ris} and \ac{isac} are emerging technologies that are the focus of growing research attention. However, their fusion into \ac{ris}-empowered \ac{isac} systems is a relatively new area of research. The high level analysis and the concrete example detailed in the previous section provides a glimpse into the potential benefits of the proper combination of these two emerging technologies, and indicates that these gains go beyond marginal extensions of known results from the more established literature on \ac{ris}-empowered wireless communications. These insights give rise to several core research directions which must be explored to further unveil the capabilities of \ac{ris}-empowered \ac{isac} and to strengthen our initial observations presented in this article. We thus conclude this survey with a discussion on some of these exciting open challenges that can serve as future research directions. 

\smallskip
{\bf Beyond Simple \ac{ris} Models:} 
The derivations and analysis of \acp{ris} for sensing and \ac{isac}, as detailed in the previous sections, adopted a simplified model for the \ac{ris} behaviour, where: $1)$ the elements of the \ac{ris} can be individually tuned to exhibit any desired phase shift; and $2)$ the effect of the \ac{ris} on the overall channel obeyed the cascaded channel model in \eqref{eqn:Cascaded}. While this simplified model is analytically tractable and facilitates derivations, it is likely to not faithfully describe the overall operation of \acp{ris} in non-purely LoS channels. In particular, as discussed in Section~\ref{sec:basics}, one can often only control the \ac{ris} elements to within a finite, possibly binary, set of configurations. Furthermore, the cascaded form is a narrowband approximation \cite{faqiri2022physfad}, while both sensing and communications signals are rarely narrowband. Thus, this model is  often violated, particularly in the presence of multipath and rich scattering environments \cite{alexandropoulos2021reconfigurable}. Finally, in wideband settings, the response of each of the \ac{ris} elements is frequency-selective exhibiting dominant coupling between its behaviour in different frequencies~\cite{katsanos2022wideband}. These characteristics indicate that, in order to fully harness the potential of \acp{ris} in empowering \ac{isac} systems, dedicated signal processing methods should be derived which are aware and geared towards more physically-compliant models of the system operation. 

\smallskip
{\bf Beyond Simple \ac{isac} Models:} The core insights drawn in the previous section, regarding the role of the \ac{snc} channel coupling in \ac{isac} systems and the ability of \acp{ris} to strengthen it, were exemplified and visualized for the simplistic \ac{isac} setting described in \eqref{motivating_model}. This motivates the exploration and formulation of signal processing techniques to exploit \acp{ris} in more involved and realistic \ac{isac} system settings. For the communications setting, such extensions include the consideration of multiple \acp{ue} and interference, fading channels, as well as modulation techniques known to be suitable for \ac{isac}, e.g., orthogonal frequency division multiplexing (OFDM), orthogonal time frequency space (OTFS) modulation, or index modulation~\cite{ma2020joint}. For the sensing functionality, relevant extensions are the incorporation of Doppler frequencies for moving targets and clutter, as well as accounting for specific desirable sensing waveforms, e.g., frequency modulated continuous wave signaling~\cite{ma2021frac}. Additional considerations which should be accounted for stem from the transmitter hardware architecture, where large-scale antenna arrays may be implemented using hybrid analog/digital and holographic MIMO architectures \cite{huang2020holographic}. All these affect the beamforming capabilities of the dual-function transmitter, and must thus be taken into account in the design of signal processing methods for \ac{ris}-empowered \ac{isac} systems.

\smallskip
{\bf Fundamental Limits of \ac{ris}-Empowered \ac{isac}:}
\ac{isac} systems are inherently characterized by the trade-off they achieve between their communications and sensing capabilities \cite{chiriyath2017radar_commun}. As these fundamental trade-offs are inherently determined by the considered models for sensing and communication, i.e., the environment, they are likely to be affected by the incorporation of \acp{ris}. This motivates the characterization of these trade-offs for \ac{ris}-aided \ac{isac}, as such studies can rigorously identify the benefits that the \ac{ris} technology brings to \ac{isac} systems in a manner which is not specific to a given transmission scheme. 

\smallskip
{\bf Mobile \ac{isac} Systems in \ac{ris}-Parameterized Settings:}
A key characteristic of \ac{isac} systems, which has not been taken into account so far is mobility. The inherent savings of \ac{isac} in power, cost, hardware, and size make it highly attractive for vehicular applications, such as self-driving cars~\cite{ma2020joint} and unmanned aerial vehicles~\cite{jing2022isac,ISAC_HRIS2022}. Furthermore, the sensing functionality often involves the tracking of moving targets, rather than their instantaneous identification. As opposed to \ac{isac} systems, \acp{ris} are likely to be static and deployed in specific known locations. However, in very recent considerations \cite{Keykhosravi2022infeasible}, RISs can be embedded in moving vehicles, thus having unknown position and orientation, rendering ISAC even more complicated.

Accounting for mobility in \ac{ris}-aided \ac{isac} brings forth a multitude of research opportunities.  For once, one can envision the location of \acp{ris} and their configuration to affect the trajectory of mobile systems, motivating a joint design for trajectories and beamforming in \ac{ris}-aided \ac{isac}. Furthermore, the need to track mobile targets over given time horizons can benefit from proper joint design of \ac{ris} reflection patterns and \ac{isac} transmission schemes. Finally, the presence of mobility affects the basic model for the communication and sensing channels, e.g.,~\eqref{motivating_model}, and must thus be accounted for in the design of signal processing schemes, even when there is no need to plan trajectories or track moving targets.

\smallskip
{\bf RIS Hardware Requirements and System Design:} 
The design of \ac{ris} hardware is a topic which is still under development to date~\cite{RIS_standardization,CE_overview_2022}. Most of the current designs are mostly geared towards the application of \acp{ris} in wireless communications, although very recently, hybrid passive-active \acp{ris} \cite{alexandropoulos2021hybrid} architectures with simultaneous reflection (thus communications) and sensing capabilities are being proposed, mainly with the objective to facilitate their efficient reconfigurability. Consequently, understanding the unique hardware requirements for designing \acp{ris} which follow from their application in \ac{isac}, can contribute to the design of future efficient \ac{ris} hardware technologies. 

The deployment and management of \acp{ris} for \ac{ris}-aided/-enabled \ac{isac} is another critical research direction.  
The former includes the design of \acp{ris} for dual function \ac{isac} signals, which may be wideband or multi-band, as well as for the identification of the level of the required configurability to achieve programmable coupling between the \ac{snc} channels. The management aspect involves investigations on the deployment of multiple \acp{ris} as well as their joint orchestration for unveiling their potential in improving the ability of mobile devices to jointly sense and communicate.    

\section{Conclusion}
\label{sec:conclusions}
This article overviewed the potential of \acp{ris} for future ISAC systems, highlighting key advantages arising from the integration of these two emerging technologies. By considering a basic MIMO ISAC system model, we showcased that joint \acl{snc} designs are mostly beneficial when the respective channels are coupled, and that \acp{ris} can facilitate the efficient control of this beneficial coupling. We discussed the main signal processing challenges resulting from the fusion of these technologies and presented several future research directions. 
 
\balance
\bibliographystyle{IEEEtran}
\bibliography{IEEEabrv,refs,references}

\end{document}

%% file: macros.tex
\def\cast{{
   \mathord{
      \hbox to 0em{
         \ooalign{
	   \smash{\hbox{$\ast$}}\crcr
	   \smash{\hskip-1pt\Large\hbox{$\circ$}} }
	 \hidewidth}
      \phantom{\bigcirc}
} }}

\def\bm#1{\mbox{\boldmath $#1$}}

\newcommand{\bds}{\begin {itemize}}
\newcommand{\eds}{\end {itemize}}
\newcommand{\bdf}{\begin{definition}}
\newcommand{\blm}{\begin{lemma}}
\newcommand{\edf}{\end{definition}}
\newcommand{\elm}{\end{lemma}}
\newcommand{\bthm}{\begin{theorem}}
\newcommand{\ethm}{\end{theorem}}
\newcommand{\bprp}{\begin{prop}}
\newcommand{\eprp}{\end{prop}}
\newcommand{\bcl}{\begin{claim}}
\newcommand{\ecl}{\end{claim}}
\newcommand{\bcr}{\begin{coro}}
\newcommand{\ecr}{\end{coro}}
\newcommand{\bquest}{\begin{question}}
\newcommand{\equest}{\end{question}}

\newcommand{\larrow}{{\larrow}}

\newcommand{\argmin}{\ensuremath{\mathrm{arg}\min}}
\newcommand{\argmax}{\ensuremath{\mathrm{arg}\max}}

\newcommand{\va}{{\ensuremath{{\mathbf{a}}}}}

\newcommand{\vc}{{\ensuremath{{\mathbf{c}}}}}

\newcommand{\vh}{{\ensuremath{{\mathbf{h}}}}}

\newcommand{\vs}{{\ensuremath{{\mathbf{s}}}}}

\newcommand{\vw}{{\ensuremath{{\mathbf{w}}}}}

\newcommand{\vx}{{\ensuremath{{\mathbf{x}}}}}

\newcommand{\mG}{{\ensuremath{\mathbf{G}}}}

\newcommand{\mI}{{\ensuremath{\mathbf{I}}}}

\newcommand{\mR}{{\ensuremath{\mathbf{R}}}}

\newcommand{\mW}{{\ensuremath{\mathbf{W}}}}

\usepackage{latexsym}

\def\IC{\mathbb C}
\def\IN{\mathbb N}
\def\IZ{\mathbb Z}
\def\IR{\mathbb R}

\def\shat{^{\mathchoice{}{}%
 {\,\,\smash{\hbox{\lower4pt\hbox{$\widehat{\null}$}}}}%
 {\,\smash{\hbox{\lower3pt\hbox{$\hat{\null}$}}}}}}

\def\bSigma{{
      \ooalign{
      \smash{\hskip.4pt\raise.4pt\hbox{$\Sigma$}}\vphantom{}\crcr
      \smash{\hskip.7pt\raise.6pt\hbox{$\Sigma$}}\vphantom{}\crcr
      \smash{\hbox{$\Sigma$}}\vphantom{$\Sigma$}}
      \vphantom{\hbox{$\Sigma$}}
      }}
\def\bTheta{{
      \ooalign{
      \smash{\hskip.5pt\raise.5pt\hbox{$\Theta$}}\vphantom{}\crcr
      \smash{\hskip.0pt\raise.1pt\hbox{$\Theta$}}\vphantom{}\crcr
      \smash{\hbox{$\Theta$}}\vphantom{$\Theta$}}
      \vphantom{\hbox{$\Theta$}}
      }}
\def\bDelta{{
      \ooalign{
      \smash{\hskip.4pt\raise.4pt\hbox{$\Delta$}}\vphantom{}\crcr
      \smash{\hskip.7pt\raise.6pt\hbox{$\Delta$}}\vphantom{}\crcr
      \smash{\hbox{$\Delta$}}\vphantom{$\Delta$}}
      \vphantom{\hbox{$\Delta$}}
      }}
\def\bLambda{{
      \ooalign{
      \smash{\hskip.5pt\raise.5pt\hbox{$\Lambda$}}\vphantom{}\crcr
      \smash{\hskip.0pt\raise.1pt\hbox{$\Lambda$}}\vphantom{}\crcr
      \smash{\hbox{$\Lambda$}}\vphantom{$\Lambda$}}
      \vphantom{\hbox{$\Lambda$}}
      }}

\makeatletter

\def\bordermatrix#1{\begingroup \m@th
  \@tempdima 8.75\p@
  \setbox\z@\vbox{%
    \def\cr{\crcr\noalign{\kern2\p@\global\let\cr\endline}}%
    \ialign{$##$\hfil\kern2\p@\kern\@tempdima&\thinspace\hfil$##$\hfil
      &&\quad\hfil$##$\hfil\crcr
      \omit\strut\hfil\crcr\noalign{\kern-\baselineskip}%
      #1\crcr\omit\strut\cr}}%
  \setbox\tw@\vbox{\unvcopy\z@\global\setbox\@ne\lastbox}%
  \setbox\tw@\hbox{\unhbox\@ne\unskip\global\setbox\@ne\lastbox}%
  \setbox\tw@\hbox{$\kern\wd\@ne\kern-\@tempdima\left[\kern-\wd\@ne
    \global\setbox\@ne\vbox{\box\@ne\kern2\p@}%
    \vcenter{\kern-\ht\@ne\unvbox\z@\kern-\baselineskip}\,\right]$}%
  \null\;\vbox{\kern\ht\@ne\box\tw@}\endgroup}
\makeatother

\makeatletter
\def\argmin{\mathop{\operator@font arg\,min}}
\def\argmax{\mathop{\operator@font arg\,max}}
\makeatother

\def\bm#1{\mbox{\boldmath $#1$}}

\newcommand{\bea}{\begin{array}}
\newcommand{\ena}{\end{array}}
\newcommand{\beq}{\begin{equation}}
\newcommand{\enq}{\end{equation}}

\newcommand{\beqa}{\begin{eqnarray}}
\newcommand{\enqa}{\end{eqnarray}}

\newcommand{\beqan}{\begin{eqnarray*}}
\newcommand{\enqan}{\end{eqnarray*}}

\newcommand{\AL}{\begin{enumerate}}
\newcommand{\ALE}{\end{enumerate}}

\def\addots{\mathinner{
    \mkern1mu\raise0pt\vbox{\kern7pt\hbox{.}}
    \mkern2mu\raise4pt\hbox{.}
    \mkern2mu\raise7pt\hbox{.}
    \mkern1mu}}

\def\sddots{\mathinner{
    \mkern.8mu\raise7pt\hbox{.}
    \mkern.8mu\raise4pt\hbox{.}
    \mkern.8mu\raise0pt\vbox{\kern7pt\hbox{.}}
    \mkern1mu}}

\def\saddots{\mathinner{
    \mkern.2mu\raise0pt\vbox{\kern7pt\hbox{.}}
    \mkern.2mu\raise4pt\hbox{.}
    \mkern.2mu\raise7pt\hbox{.}
    \mkern1mu}}

\newcommand{\bw}{{\ensuremath{\mathbf{w}}}}

\def\sqplus{\mathbin{
	{\ooalign{\hfil\raise.3ex\hbox{\scriptsize
	+}\hfil\crcr\mathhexbox274\crcr\mathhexbox275}}
	}} 
\def\sqminus{\mathbin{
	{\ooalign{\hfil\raise.3ex\hbox{\scriptsize
	--}\hfil\crcr\mathhexbox274\crcr\mathhexbox275}}
	}}

\def\IC{{
   \mathord{
      \hbox to 0em{
	 \hskip-4pt
         \ooalign{
	   \smash{\hskip1.9pt\raise2.6pt\hbox{$\scriptscriptstyle |$}}\crcr
	   \smash{\hbox{\rm\sf C}} }
	 \hidewidth}
      \phantom{\hbox{\rm\sf C}}
} }}
\def\IN{
    {\ooalign{
   \smash{\hskip2.2pt\raise1.5pt\hbox{$\scriptscriptstyle |$}}\vphantom{}\crcr
   \hbox{\sf N}
	}}
	} 
\def\IZ{
    {\ooalign{
   \smash{\hskip1.9pt\raise0pt\hbox{$\sf Z$}}\vphantom{}\crcr
   \hbox{\sf Z}
	}}
	} 
\def\IR{
    {\ooalign{
   \smash{\hskip2.2pt\raise1.5pt\hbox{$\scriptscriptstyle |$}}\vphantom{}\crcr
   \smash{\hskip2.2pt\raise3.3pt\hbox{$\scriptscriptstyle |$}}\vphantom{}\crcr
   \hbox{\sf R}
	}}
	} 

\DeclareMathAlphabet{\mathcmb}{OT1}{cmr}{b}{n}

\def\bSigma{\ensuremath{\mathcmb{\Sigma}}}
\def\bLambda{\ensuremath{\mathcmb{\Lambda}}}

\def\bTheta{\ensuremath{\mathcmb{\Theta}}}

\newcommand{\SI}{\begin{indlist}}
\newcommand{\EI}{\end{indlist}}

\newcommand{\DL}{\begin{dashlist}}
\newcommand{\DLE}{\end{dashlist}}

\makeatletter
\def\setboxz@h{\setbox\z@\hbox}
\def\wdz@{\wd\z@}
\def\boxz@{\box\z@}
\def\underset#1#2{\binrel@{#2}%
  \binrel@@{\mathop{\kern\z@#2}\limits_{#1}}}
\def\binrel@#1{\begingroup
  \setboxz@h{\thinmuskip0mu
    \medmuskip\m@ne mu\thickmuskip\@ne mu
    \setbox\tw@\hbox{$#1\m@th$}\kern-\wd\tw@
    ${}#1{}\m@th$}%
  \edef\@tempa{\endgroup\let\noexpand\binrel@@
    \ifdim\wdz@<\z@ \mathbin
    \else\ifdim\wdz@>\z@ \mathrel
    \else \relax\fi\fi}%
  \@tempa
}
\let\binrel@@\relax%
\makeatother


%% file: ris_isac.bbl
\begin{thebibliography}{10}
\providecommand{\url}[1]{#1}
\csname url@samestyle\endcsname
\providecommand{\newblock}{\relax}
\providecommand{\bibinfo}[2]{#2}
\providecommand{\BIBentrySTDinterwordspacing}{\spaceskip=0pt\relax}
\providecommand{\BIBentryALTinterwordstretchfactor}{4}
\providecommand{\BIBentryALTinterwordspacing}{\spaceskip=\fontdimen2\font plus
\BIBentryALTinterwordstretchfactor\fontdimen3\font minus
  \fontdimen4\font\relax}
\providecommand{\BIBforeignlanguage}[2]{{%
\expandafter\ifx\csname l@#1\endcsname\relax
\typeout{** WARNING: IEEEtran.bst: No hyphenation pattern has been}%
\typeout{** loaded for the language `#1'. Using the pattern for}%
\typeout{** the default language instead.}%
\else
\language=\csname l@#1\endcsname
\fi
#2}}
\providecommand{\BIBdecl}{\relax}
\BIBdecl

\bibitem{RIS_standardization}
R.~Liu, G.~C. Alexandropoulos, Q.~Wu, M.~Jian, and Y.~Liu, ``How can
  reconfigurable intelligent surfaces drive {5G}-advanced wireless networks:
  {A} standardization perspective,'' in \emph{Proc. IEEE/CIC Int. Conf. Commun.
  in China}, Foshan, China, Aug. 2022.

\bibitem{huang2019reconfigurable}
C.~Huang, A.~Zappone, G.~C. Alexandropoulos, M.~Debbah, and C.~Yuen,
  ``Reconfigurable intelligent surfaces for energy efficiency in wireless
  communication,'' \emph{{IEEE} Trans. Wireless Commun.}, vol.~18, no.~8, pp.
  4157--4170, 2019.

\bibitem{di2019smart}
M.~Di~Renzo, M.~Debbah, D.-T. Phan-Huy, A.~Zappone, M.-S. Alouini, C.~Yuen,
  V.~Sciancalepore, G.~C. Alexandropoulos, J.~Hoydis, H.~Gacanin, J.~de~Rosny,
  A.~Bounceu, G.~Lerosey, and M.~Fink, ``Smart radio environments empowered by
  reconfigurable {AI} meta-surfaces: an idea whose time has come,''
  \emph{EURASIP J. Wireless Commun. Net.}, vol. 2019, no.~1, pp. 1--20, 2019.

\bibitem{risTUTORIAL2020}
Q.~Wu, S.~Zhang, B.~Zheng, C.~You, and R.~Zhang, ``Intelligent reflecting
  surface aided wireless communications: {A} tutorial,'' \emph{IEEE Trans.
  Commun.}, vol.~69, no.~5, pp. 3313--3351, May 2021.

\bibitem{RISE6G_COMMAG}
E.~Calvanese~Strinati, G.~C. Alexandropoulos, H.~Wymeersch, B.~Denis,
  V.~Sciancalepore, R.~D'Errico, A.~Clemente, D.-T. Phan-Huy, E.~D. Carvalho,
  and P.~Popovski, ``Reconfigurable, intelligent, and sustainable wireless
  environments for {6G} smart connectivity,'' \emph{IEEE Commun. Mag.},
  vol.~59, no.~10, pp. 99--105, 2021.

\bibitem{chowdhury20206g}
M.~Z. Chowdhury, M.~Shahjalal, S.~Ahmed, and Y.~M. Jang, ``{6G} wireless
  communication systems: Applications, requirements, technologies, challenges,
  and research directions,'' \emph{IEEE Open Journal of the Communications
  Society}, vol.~1, pp. 957--975, 2020.

\bibitem{WavePropTCCN}
G.~C. Alexandropoulos, G.~Lerosey, M.~Debbah, and M.~Fink, ``Reconfigurable
  intelligent surfaces and metamaterials: {T}he potential of wave propagation
  control for {6G} wireless communications,'' \emph{IEEE ComSoc TCCN
  Newslett.}, vol.~6, no.~1, pp. 25--37, Jun. 2020.

\bibitem{CE_overview_2022}
M.~Jian, G.~C. Alexandropoulos, E.~Basar, C.~Huang, R.~Liu, Y.~Liu, and
  C.~Yuen, ``Reconfigurable intelligent surfaces for wireless communications:
  Overview of hardware designs, channel models, and estimation techniques,''
  \emph{Intel. Converged Net.}, vol.~3, no.~1, pp. 1--32, 2022.

\bibitem{liu2021reconfigurable}
Y.~Liu, X.~Liu, X.~Mu, T.~Hou, J.~Xu, M.~Di~Renzo, and N.~Al-Dhahir,
  ``Reconfigurable intelligent surfaces: Principles and opportunities,''
  \emph{{IEEE} Commun. Surveys Tuts.}, vol.~23, no.~3, pp. 1546--1577, 2021.

\bibitem{RIS_6G_tutorial}
J.~Xu, C.~Yuen, C.~Huang, N.~U. Hassan, G.~C. Alexandropoulos, M.~Di~Renzo, and
  M.~Debbah, ``Reconfiguring wireless environment via intelligent surfaces for
  {6G: R}eflection, modulation, and security,'' \emph{Science China Inf.
  Sciences}, to appear, 2022.

\bibitem{Samsung}
``The next hyper- {C}onnected experience for all,'' White Paper, Samsung 6G
  Vision, Jun. 2020.

\bibitem{bjornson2021reconfigurable}
E.~Bj{\"o}rnson, H.~Wymeersch, B.~Matthiesen, P.~Popovski, L.~Sanguinetti, and
  E.~de~Carvalho, ``Reconfigurable intelligent surfaces: A signal processing
  perspective with wireless applications,'' \emph{{IEEE} Signal Process. Mag.},
  vol.~39, no.~2, pp. 135--158, 2022.

\bibitem{Keykhosravi2022infeasible}
K.~Keykhosravi, B.~Denis, G.~C. Alexandropoulos, Z.~S. He, A.~Albanese,
  V.~Sciancalepore, and H.~Wymeersch, ``Leveraging {RIS}-enabled smart signal
  propagation for solving infeasible localization problems,'' \emph{arXiv
  preprint arXiv:2204.11538}, 2022.

\bibitem{elzanaty2021reconfigurable}
A.~Elzanaty, A.~Guerra, F.~Guidi, and M.-S. Alouini, ``Reconfigurable
  intelligent surfaces for localization: Position and orientation error
  bounds,'' \emph{{IEEE} Trans. Signal Process.}, vol.~69, pp. 5386--5402,
  2021.

\bibitem{ma2020joint}
D.~Ma, N.~Shlezinger, T.~Huang, Y.~Liu, and Y.~C. Eldar, ``Joint
  radar-communication strategies for autonomous vehicles: Combining two key
  automotive technologies,'' \emph{{IEEE} Signal Process. Mag.}, vol.~37,
  no.~4, pp. 85--97, 2020.

\bibitem{Liu_ISAC}
F.~Liu, C.~Masouros, A.~P. Petropulu, H.~Griffiths, and L.~Hanzos, ``Joint
  radar and communication design: {A}pplications, state-of-the-art, and the
  road ahead,'' \emph{IEEE Trans. Commun.}, vol.~68, no.~6, pp. 3834--3862,
  June 2020.

\bibitem{liu2021integrated}
F.~Liu, Y.~Cui, C.~Masouros, J.~Xu, T.~X. Han, Y.~C. Eldar, and S.~Buzzi,
  ``Integrated sensing and communications: Towards dual-functional wireless
  networks for {6G} and beyond,'' \emph{{IEEE} J. Sel. Areas Commun.}, vol.~40,
  no.~6, pp. 1728--1767, 2022.

\bibitem{Zhang_ISAC}
J.~A. Zhang, F.~Liu, C.~Masouros, R.~W. Heath, Z.~Feng, L.~Zheng, and
  A.~Petropulu, ``An overview of signal processing techniques for joint
  communication and radar sensing,'' \emph{IEEE J. Sel. Topics Signal
  Process.}, vol.~15, no.~6, pp. 1295--1315, 2021.

\bibitem{Cui_ISAC}
Y.~Cui, F.~Liu, X.~Jing, and J.~Mu, ``Integrating sensing and communications
  for ubiquitous iot: {A}pplications, trends, and challenges,'' \emph{IEEE
  Network}, vol.~35, no.~5, pp. 158--167, 2021.

\bibitem{buzzi2021foundations}
S.~Buzzi, E.~Grossi, M.~Lops, and L.~Venturino, ``Foundations of {MIMO} radar
  detection aided by reconfigurable intelligent surfaces,'' \emph{{IEEE} Trans.
  Signal Process.}, vol.~70, pp. 1749--1763, 2022.

\bibitem{alexandropoulos2021hybrid}
G.~C. Alexandropoulos, N.~Shlezinger, I.~Alamzadeh, M.~F. Imani, H.~Zhang, and
  Y.~C. Eldar, ``Hybrid reconfigurable intelligent metasurfaces: Enabling
  simultaneous tunable reflections and sensing for {6G} wireless
  communications,'' \emph{arXiv preprint arXiv:2104.04690}, 2021.

\bibitem{shlezinger2020dynamic}
N.~Shlezinger, G.~C. Alexandropoulos, M.~F. Imani, Y.~C. Eldar, and D.~R.
  Smith, ``Dynamic metasurface antennas for {6G} extreme massive {MIMO}
  communications,'' \emph{{IEEE} Wireless Commun.}, vol.~28, no.~2, pp.
  106--113, 2021.

\bibitem{huang2020holographic}
C.~Huang, S.~Hu, G.~C. Alexandropoulos, A.~Zappone, C.~Yuen, R.~Zhang,
  M.~Di~Renzo, and M.~Debbah, ``Holographic {MIMO} surfaces for {6G} wireless
  networks: Opportunities, challenges, and trends,'' \emph{{IEEE} Wireless
  Commun.}, vol.~27, no.~5, pp. 118--125, 2020.

\bibitem{mu2021simultaneously}
X.~Mu, Y.~Liu, L.~Guo, J.~Lin, and R.~Schober, ``Simultaneously transmitting
  and reflecting {(STAR) RIS} aided wireless communications,'' \emph{{IEEE}
  Trans. Wireless Commun.}, vol.~21, no.~5, pp. 3083--3098, 2021.

\bibitem{Alamzadeh2021ris}
I.~Alamzadeh, G.~C. Alexandropoulos, N.~Shlezinger, and M.~F. Imani, ``{A
  reconfigurable intelligent surface with integrated sensing capability},''
  \emph{Scientific Reports}, vol.~11, no.~1, p. 20737, 2021.

\bibitem{pan2021reconfigurable}
C.~Pan, H.~Ren, K.~Wang, J.~F. Kolb, M.~Elkashlan, M.~Chen, M.~Di~Renzo,
  Y.~Hao, J.~Wang, A.~L. Swindlehurst, X.~You, and L.~Hanzo, ``Reconfigurable
  intelligent surfaces for {6G} systems: Principles, applications, and research
  directions,'' \emph{{IEEE} Commun. Mag.}, vol.~59, no.~6, pp. 14--20, 2021.

\bibitem{chen2020angle}
W.~Chen, L.~Bai, W.~Tang, S.~Jin, W.~X. Jiang, and T.~J. Cui, ``Angle-dependent
  phase shifter model for reconfigurable intelligent surfaces: Does the
  angle-reciprocity hold?'' \emph{{IEEE} Commun. Lett.}, vol.~24, no.~9, pp.
  2060--2064, 2020.

\bibitem{katsanos2022wideband}
K.~D. Katsanos, N.~Shlezinger, M.~F. Imani, and G.~C. Alexandropoulos,
  ``Wideband multi-user {MIMO} communications with frequency selective {RISs}:
  Element response modeling and sum-rate maximization,'' in \emph{IEEE
  International Conference on Communications (ICC)}, 2022.

\bibitem{faqiri2022physfad}
R.~Faqiri, C.~Saigre-Tardif, G.~C. Alexandropoulos, N.~Shlezinger, M.~F. Imani,
  and P.~del Hougne, ``Phys{F}ad: Physics-based end-to-end channel modeling of
  {RIS}-parametrized environments with adjustable fading,'' \emph{{IEEE} Trans.
  Wireless Commun.}, 2022, early access.

\bibitem{georgeRISpls2021}
G.~C. Alexandropoulos, K.~Katsanos, M.~Wen, and D.~B. da~Costa, ``Safeguarding
  {MIMO} communications with reconfigurable metasurfaces and artificial
  noise,'' in \emph{Proc. IEEE ICC}, Montreal, Canada, jun. 2021.

\bibitem{wu2019beamforming}
Q.~Wu and R.~Zhang, ``Beamforming optimization for wireless network aided by
  intelligent reflecting surface with discrete phase shifts,'' \emph{{IEEE}
  Trans. Commun.}, vol.~68, no.~3, pp. 1838--1851, 2019.

\bibitem{basar2020reconfigurable}
E.~Basar, ``Reconfigurable intelligent surface-based index modulation: A new
  beyond {MIMO} paradigm for {6G},'' \emph{{IEEE} Trans. Commun.}, vol.~68,
  no.~5, pp. 3187--3196, 2020.

\bibitem{reflection_pattern_modulation2021}
S.~Lin, B.~Zheng, G.~C. Alexandropoulos, M.~Wen, M.~Di~Renzo, and F.~Chen,
  ``Reconfigurable intelligent surfaces with reflection pattern modulation:
  {B}eamforming design, channel estimation, and achievable rate analysis,''
  \emph{IEEE Trans. Wireless Commun.}, vol.~20, no.~2, pp. 741--754, 2021.

\bibitem{yang2021reconfigurable}
H.~Yang, X.~Yuan, J.~Fang, and Y.-C. Liang, ``Reconfigurable intelligent
  surface aided constant-envelope wireless power transfer,'' \emph{{IEEE}
  Trans. Signal Process.}, vol.~69, pp. 1347--1361, 2021.

\bibitem{george_henk_ICC2021}
Z.~Abu-Shaban, K.~Keykhosravi, M.~F. Keskin, G.~C. Alexandropoulos,
  G.~Seco-Granados, and H.~Wymeersch, ``Near-field localization with a
  reconfigurable intelligent surface acting as lens,'' in \emph{IEEE
  International Conference on Communications (ICC)}, 2021.

\bibitem{Kimmo_Popovski_2021}
C.~J. Vaca-Rubio, P.~Ramirez-Espinosa, K.~Kansanen, Z.-H. Tan, E.~De~Carvalho,
  and P.~Popovski, ``Assessing wireless sensing potential with large
  intelligent surfaces,'' \emph{IEEE Open Journal of the Communications
  Society}, vol.~2, pp. 934--947, 2021.

\bibitem{hu2020reconfigurable}
J.~Hu, H.~Zhang, B.~Di, L.~Li, K.~Bian, L.~Song, Y.~Li, Z.~Han, and H.~V. Poor,
  ``Reconfigurable intelligent surface based {RF} sensing: Design,
  optimization, and implementation,'' \emph{{IEEE} J. Sel. Areas Commun.},
  vol.~38, no.~11, pp. 2700--2716, 2020.

\bibitem{he2022risassisted}
Y.~He, Y.~Cai, H.~Mao, and G.~Yu, ``{RIS}-assisted communication radar
  coexistence: Joint beamforming design and analysis,'' \emph{{IEEE} J. Sel.
  Areas Commun.}, vol.~40, no.~7, pp. 2131--2145, 2022.

\bibitem{sankar2021joint}
R.~Prasobh~Sankar, B.~Deepak, and S.~P. Chepuri, ``Joint communication and
  radar sensing with reconfigurable intelligent surfaces,'' in \emph{IEEE
  International Workshop on Signal Processing Advances in Wireless
  Communications (SPAWC)}, 2021, pp. 471--475.

\bibitem{sankar2022beamforming}
R.~Prasobh~Sankar, S.~P. Chepuri, and Y.~C. Eldar, ``Beamforming in integrated
  sensing and communication systems with reconfigurable intelligent surfaces,''
  \emph{arXiv preprint arXiv:2206.07679}, 2022.

\bibitem{sankar2022beamforming2}
R.~Prasobh~Sankar and S.~P. Chepuri, ``Beamforming in hybrid {RIS}-assisted
  integrated sensing and communication systems,'' \emph{arXiv preprint
  arXiv:2203.05902}, 2022.

\bibitem{chiriyath2017radar_commun}
A.~R. Chiriyath, B.~Paul, and D.~W. Bliss, ``Radar-communications convergence:
  Coexistence, cooperation, and co-design,'' \emph{{IEEE} Trans. on Cogn.
  Commun. Netw.}, vol.~3, no.~1, pp. 1--12, Mar. 2017.

\bibitem{Abeywickrama_2020}
S.~Abeywickrama, R.~Zhang, Q.~Wu, and C.~Yue, ``Intelligent reflecting surface:
  {P}ractical phase shift model and beamforming optimization,'' \emph{IEEE
  Trans. Commun.}, vol.~68, no.~9, pp. 5849--5863, 2020.

\bibitem{van2004optimum}
H.~L. Van~Trees, \emph{Optimum array processing: Part IV of detection,
  estimation, and modulation theory}.\hskip 1em plus 0.5em minus 0.4em\relax
  John Wiley \& Sons, 2004.

\bibitem{liu_CRB}
F.~Liu, Y.-F. Liu, A.~Li, C.~Masouros, and Y.~C. Eldar, ``Cram\'er-{R}ao bound
  optimization for joint radar-communication beamforming,'' \emph{{IEEE} Trans.
  Signal Process.}, vol.~70, pp. 240--253, 2022.

\bibitem{Meng_arxiv}
X.~Meng, F.~Liu, S.~Lu, S.~P. Chepuri, and C.~Masouros, ``{RIS}-assisted
  integrated sensing and communications: {A} subspace rotation approach,''
  \emph{arXiv preprint arXiv:2210.13987}, 2022.

\bibitem{alexandropoulos2021reconfigurable}
G.~C. Alexandropoulos, N.~Shlezinger, and P.~del Hougne, ``Reconfigurable
  intelligent surfaces for rich scattering wireless communications: Recent
  experiments, challenges, and opportunities,'' \emph{{IEEE} Commun. Mag.},
  vol.~59, no.~6, pp. 28--34, 2021.

\bibitem{ma2021frac}
D.~Ma, N.~Shlezinger, T.~Huang, Y.~Liu, and Y.~C. Eldar, ``{FRaC}: {FMCW}-based
  joint radar-communications system via index modulation,'' \emph{{IEEE} J.
  Sel. Topics Signal Process.}, vol.~15, no.~6, pp. 1348--1364, 2021.

\bibitem{jing2022isac}
X.~Jing, F.~Liu, C.~Masouros, and Y.~Zeng, ``{ISAC} from the sky: {UAV}
  trajectory design for joint communication and target localization,''
  \emph{arXiv preprint arXiv:2207.02904}, 2022.

\bibitem{ISAC_HRIS2022}
J.~He, A.~Fakhreddine, and G.~C. Alexandropoulos, ``Joint channel and direction
  estimation for ground-to-{UAV} communications enabled by a simultaneous
  reflecting and sensing {RIS},'' \emph{arXiv preprint arXiv:2210.15238}, 2022.

\end{thebibliography}
